\documentclass[twocolumn,floatfix,amssymb,amsmath,secnumarabic,nofootinbib, nobalancelastpage]{revtex4-1}    

\usepackage[pdftex]{graphicx} 
\usepackage{dcolumn}  
\usepackage{bm}       
\usepackage[usenames,dvipsnames]{color}														
\definecolor{URLCOL}{rgb}{0,0.52,0.83} 
\definecolor{LINKCOL}{rgb}{0.05,0.5,0} 
\definecolor{CITECOL}{rgb}{0.25,0,0.48} 
\usepackage{epstopdf}
\usepackage[pdftex,bookmarks,breaklinks,bookmarksopen,bookmarksnumbered,colorlinks,
		    linkcolor=LINKCOL,linktocpage,citecolor=CITECOL,urlcolor=URLCOL,
			pdfpagemode=UseOutline,pdftex,pagebackref]{hyperref}
\usepackage{datetime}

\def\preprintlink{ \href{http://dft.uci.edu + PAPER REF}{title of paper} }
\def\preprinttext{~}

\usepackage{fancyhdr}
\makeatletter
\fancypagestyle{titlepage}
{
	\lhead{\textsc{\preprinttext\ ~ \href{http://dft.uci.edu/publications.php}{~}}}
	\chead{}
	\rhead{\preprintlink}
	\lfoot{}
}
\makeatother
\def\preprintlink{ 
	\href{http://dft.uci.edu}
        {
~}
	}

\usepackage{graphicx}
\usepackage{amsmath}
\usepackage{physics} 
\usepackage{mathrsfs} 
\usepackage{subfigure} 
\usepackage{dcolumn} 
\usepackage{tabularx}
\usepackage{bm}
\usepackage{threeparttable}
\usepackage[normalem]{ulem}
\usepackage{array,multirow}
\usepackage{appendix}
\usepackage{epstopdf}
\usepackage{booktabs}
\usepackage{natbib}
\usepackage{feynmp}
\usepackage{threeparttable}
\usepackage{CJKutf8}

\definecolor{TITLECOL}{rgb}{0.1,0.2,0.7} 
\definecolor{PCOL}{rgb}{0.5,0.06,0.01} 
\definecolor{CHAPCOL}{rgb}{0,0.48,0} 
\definecolor{SECOL}{rgb}{0.1,0.2,0.7} 
\definecolor{CONTENTSCOL}{rgb}{0.1,0.2,0.7} 
\definecolor{SSECOL}{rgb}{0.25,0,0.48} 
\definecolor{SSSECOL}{rgb}{0.2,0.08,0.53} 
\definecolor{SHDCOL}{rgb}{0.4,0,0} 
\definecolor{ITMCOL}{rgb}{0.4,0,0} 
\definecolor{EXCOL}{rgb}{0,0.47,0.01} 
\definecolor{DEFCOL}{rgb}{0,0.42,0.01} 

\def\coloredtitle#1{\title{\textcolor{TITLECOL}{#1}}} 

\definecolor{URLCOL}{rgb}{0,0.17,0.43} 
\definecolor{LINKCOL}{rgb}{0.05,0.4,0} 
\definecolor{CITECOL}{rgb}{0.35,0,0.48} 

\definecolor{ngreen}{rgb}{0,0.48,0}

\def\sec#1{\section{\textcolor{SECOL}{#1}}}
\def\ssec#1{\subsection{\textcolor{SSECOL}{#1}}}

\def\sectable#1{
\addcontentsline{toc}{subsection}{~~Table: \textcolor{SSECOL}{#1}}
\begin{table}[h]
\caption{\bf \textcolor{SSECOL}{#1}}
}

\def\bea{\begin{eqnarray}}
\def\eea{\end{eqnarray}}
\def\ben{\begin{equation}}
\def\een{\end{equation}}
\def\benu{\begin{enumerate}}
\def\enu{\end{enumerate}}

\def\bei{\begin{itemize}}
\def\eei{\end{itemize}}
\def\beit{\begin{itemize}}
\def\eit{\end{itemize}}
\def\benu{\begin{enumerate}}
\def\enu{\end{enumerate}}

\def\sss{\scriptscriptstyle\rm}

\def\g{_\gamma}

\def\1var{(\bx_1...\bx\N)}

\def\bx{{x}}

\def\bj{{\bf j}}

\def\N{_{\sss N}}

\def\sph_int{ {\int d^3 r}}

\definecolor{SPECOL}{rgb}{0,0.47,0.01}
\definecolor{QUOCOL}{rgb}{0,0,0.2}
\definecolor{SHDCOLb}{rgb}{0.69,0.4,0.1}

\definecolor{SPEQ}{rgb}{0.01,0.4,0.05} %

\definecolor{SPEQv}{rgb}{0.45,0.05,0.45} %

\definecolor{SPEQb}{rgb}{0.01,0.1,0.65} %

\definecolor{SPEQr}{rgb}{0.57,0.05,0.1} %

\def\sec#1{\section{\textcolor{SECOL}{#1}}}
\def\ssec#1{\subsection{\textcolor{SSECOL}{#1}}}

\def\bay{\begin{array}}
\def\eay{\end{array}}
\def\bit{\begin{itemize}}
\def\beit{\begin{itemize}}
\def\eit{\end{itemize}}

\def\ln{\text{ln} }

\def\floor{\text{floor} }

\def\dd{~ \rotatebox{320}{\hspace{-5pt}\vbox to 5 pt {\hspace{-5pt} \hbox to 5pt {$\cdots$}}}\!\! }

\graphicspath{{Figures/}}
\begin{document}


\sf 
\coloredtitle{Uncommonly accurate energies for the general quartic oscillator}
\author{\color{CITECOL} Pavel Okun, Kieron Burke}
\affiliation{Department of Chemistry,
	University of California, Irvine, CA 92697,  USA}
\affiliation{Departments of Physics and Astronomy and of Chemistry}
\date{\today}
\begin{abstract}
Recent advances in the asymptotic analysis of energy levels of potentials
produce relative errors in eigenvalue sums of order $10^{-34}$, 
but few non-trivial potentials have been solved numerically to such
accuracy. We solve the general quartic potential (arbitrary linear combination of $x^2$ and $x^4$ )
beyond this level of accuracy using a basis of several hundred oscillator states. 
We list the lowest 20 eigenvalues for 9 such
potentials. We confirm the known asymptotic expansion for the levels of the pure quartic
oscillator, and extract the next 2 terms in the asymptotic expansion. We give analytic formulas
for expansion in up to 3 even basis states. We confirm the virial theorem for the various energy
components to similar accuracy. The sextic oscillator levels are also given. These benchmark results should be useful for extreme tests of approximations in several areas of chemical physics and beyond.
\end{abstract}


\maketitle
\def\floor#1{{\lfloor}#1{\rfloor}}
\def\sm#1{{\langle}#1{\rangle}}
\def\dis{_{disc}}
\newcommand{\Z}{\mathbb{Z}}
\newcommand{\R}{\mathbb{R}}
\def\w{^{(0)}}
\def\w{^{\rm WKB}}
\def\II{^{\rm II}}
\def\hd#1{\noindent{\bf\textcolor{red} {#1:}}}
\def\hb#1{\noindent{\bf\textcolor{blue} {#1:}}}
\def\eps{\epsilon}
\def\ew{\epsilon\w}
\def\ej{\epsilon_j}
\def\upet{^{(\eta)}}
\def\ejeta{\ej\upet}
\def\tjeta{\tj\upet}
\def\bej{{\bar \epsilon}_j}
\def\ewj{\epsilon\w_j}
\def\tj{t_j}
\def\vj{v_j}
\def\F{_{\sss F}}
\def\xt{x_{\sss T}}
\def\sc{^{\rm sc}}
\def\al{\alpha}
\def\ae{\al_e}
\def\bj{\bar j}
\def\bz{\bar\zeta}
\def\eq#1{Eq.\, (\ref{#1})}
\def\cN{{\cal N}}

\graphicspath{{./F/}}
\def\lam{\lambda}
\def\G{\Gamma}
\def\g{\gamma}
\def\eps{\epsilon}
\def\om{\omega}
\def\D{\Delta}
\def\d{\delta}
\def\a{\alpha}
\def\t{\theta}
\sec{Introduction}

Since the early days of quantum mechanics, potentials with 
analytic solutions have played a crucial role in providing both
insight into more complex problems, and benchmarks for more
general quantum solution methods \cite{S26,RM32}.  The quartic oscillator
is iconic in being a simple potential without a built-in
length scale which does not have a simple analytic solution 
\cite{BO99,R70,B19,W97,KW97}.
The general quartic oscillator (adding both quadratic and
linear terms) is not scale-invariant, and has been studied in
many different contexts in physics \cite{JNG18,DP97,BW72}.  In particular, the
Mexican hat shape of symmetric double wells is a paradigm
of simple symmetry breaking \cite{JNG18,MPKD18}.   

In chemical physics, the double well provides important tests
of theories of tunneling in quantum nuclear dynamics of liquids \cite{JNG18,MPKD18,FS11}.
In particle physics, it is a prototype of symmetry breaking,
such as occurs in simple field theories \cite{DGKS93,CKS94}.  In mathematical
physics, it is a simple case to test and explore asymptotic
approximations \cite{ABS19}.
Asymptotic analysis, especially hyperasymptotics,
can yield exquisitely accurate approximations \cite{C09,BH93,BH90,BM72}.
In the past, many developments and tests of these methods have
been applied to scale invariant potentials \cite{B20b,B20,BB20}, but the
general quartic oscillator provides opportunities to look
at more complex cases.

Recent work on one-dimensional potentials \cite{B20,BB20,B20b} has established a deep explicit
connection between the gradient expansion of density functional theory
and asymptotic expansions in powers of $\hbar$ \cite{CLEB10}.  In one case
fractional errors were below the picoyocto range, i.e., of order 10$^{-33}$ \cite{BB20}.
To further develop and test methods in this area, there is a need
for benchmark calculations of this level of accuracy for
non-trivial potentials.  This exceeds even quadruple precision on
standard computers, rendering standard numerical algorithms, even pushed
to their convergence limits, difficult to apply.  There is also a new area of application:  The breaking of symmetry
is a simple prototype of a bond breaking, in which electrons
localize in two separate wells \cite{CMYb08}.   Such bond breaking is very difficult
to model with standard semilocal density functionals, and their failure
has been traced back to the change in asymptotic expansions in going from
one well to two\cite{B20}.  In some simple situations, benchmark electronic structure calculations have been performed to this level of accuracy (or higher) for systems with a few electrons \cite{NN07}.  But the purpose of the present study (and many previous ones) is to explore the underlying principles behind asymptotic (and other) approximation schemes, so as to improve the accuracy of less expensive quantum solvers, such as density functional theory, which can then be applied to much larger systems.  The benchmark data here provides a quick reference for those exploring basic questions with analytic one-dimensional models.

\begin{figure}[htb]
\includegraphics[width=0.8\columnwidth]{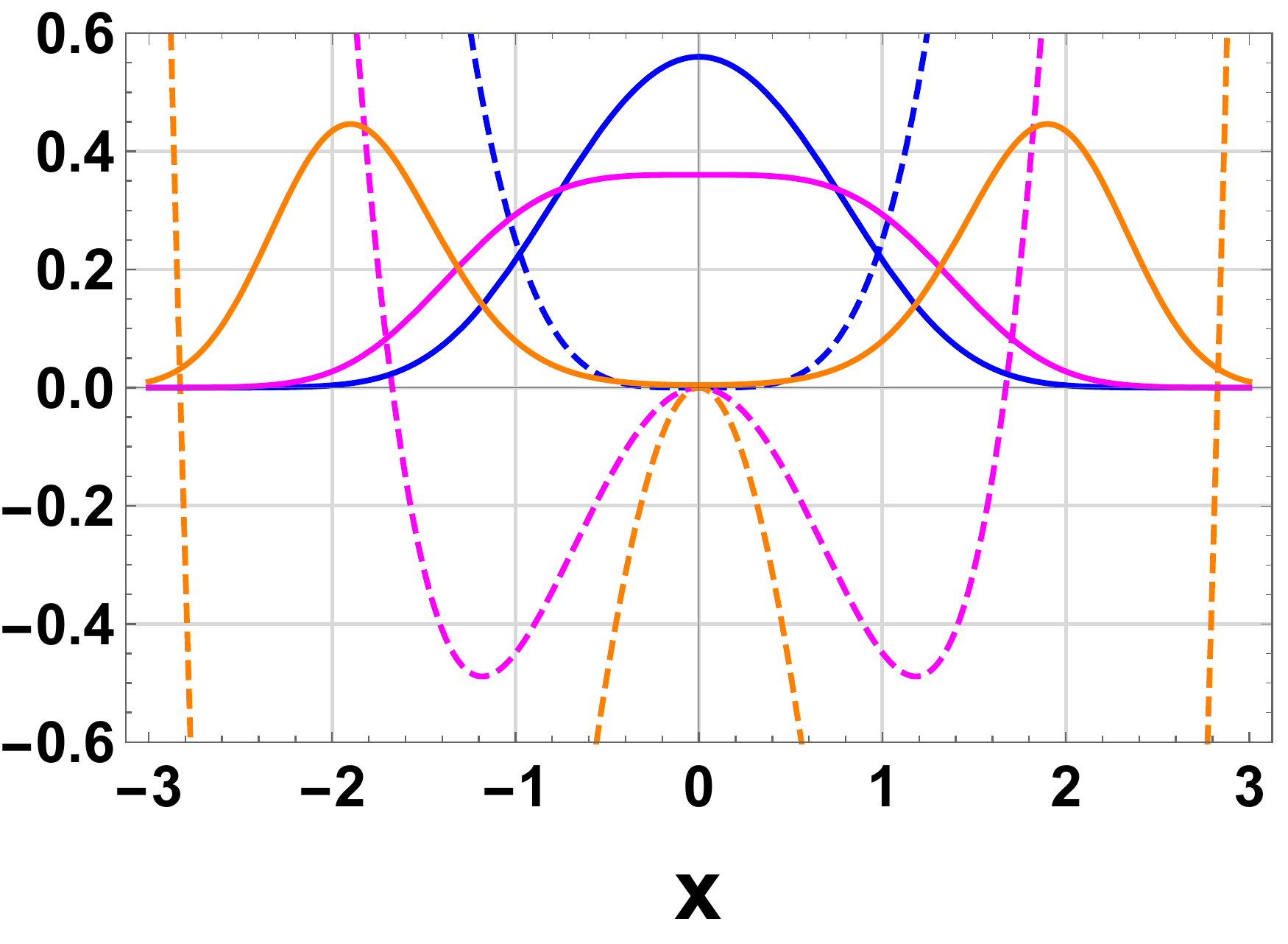}
\caption{Ground state densities (solid) and potentials (dashed).  Legend: $\lam = 0$ (blue), $\lam = \lam_c$ (magenta), $\lam = 4$ (orange).}
\label{fig:Fdens}
\end{figure}

In order to generate such benchmarks and
as a simple example, we consider the generalized quartic oscillator potential:
\begin{equation}
\label{vlambda}
v_\lambda(x) = \frac{x^4}{4} - \lambda \frac{x^2}{2},
\end{equation}
where $\lambda$ is a real number, either positive or negative.
For $\lambda=0$, this is a pure quartic oscillator, which has been the
subject of many investigations.  In this paper we will present the pure
quartic oscillator energies for more states and to more digits
than previously computed in Refs. \cite{B19,BO99,R70}.  We also numerically examine the WKB series for the quartic oscillator closely following Bender \& Orszag's book \cite{BO99}.  Previous investigations of the WKB approximation of the quartic oscillator can be found in Refs. \cite{V83,DP97,BW69,BW68,HM75}.  We examine the variation of the energy with $\lam$ and the effect of a linear term as in Ref. \cite{DP97}.  Our exact energies can be used as inputs to test the semiclassical analysis of Ref. \cite{B14}.  Other methods of estimating quartic oscillator energies are described in Refs. \cite{BO99,HM75,R70,PM20,V94,V99}.  The exact solution of the quartic oscillator was studied in Refs. \cite{W97,KW97}.
For $\lambda < 0$,
the minimum is always at $x=0$, with vibrational frequency ${\sqrt{|\lambda|}}$.
For $\lambda > 0$, the
most interesting case, two distinct wells appear, with minima at $\pm
{\sqrt{\lambda}}$, and frequency ${\sqrt{2\lambda}}$. Fig. \ref{fig:Fdens} illustrates some results, showing the density of the ground state
and the well for three values of $\lambda$: 0, $\lambda_c$ (the critical value of $\lambda$ at which the ground state energy is zero), and 4.  
The first is similar in shape to a harmonic oscillator, but with steeper walls,
and the density decays more rapidly.  The second is particularly flat, as the
energy is exactly zero.  The third is a typical double-well structure, with
two well-localized densities on each side, and a small 'overlap' at $x=0$.  Thus 
there is a transition from one well to two, and
simple symmetry breaking.  Following the behavior of asymptotic expansions
with the variation of $\lambda$ is a toy problem relevant to many fields \cite{ABS19}.

In this paper, we show how to calculate extremely accurate results for
these potentials using a symbolic manipulation code, such as Mathematica,
where manipulations can be performed with an arbitrary number of digits.
We summarize results in the main text, and provide some analysis of
various regimes.  In the supplementary information, we give many tables
of results to many digits of accuracy.

\sec{Motivation}
How can energy calculations to 40 decimal places possibly be of practical use?
Modern density functional calculations use approximations that have errors
larger than 1 kcal/mol, which is of order $10^{-5}$ of the total energy of a
Ne atom, say.  So even 1000 heavy atoms need only 9 digits of accuracy.
However, the fundamental approximation behind almost all modern density
functional approximations is the gradient expansion. 
Recent work \cite{B20,BB20,B20b}
has shown a direct, explicit connection between that expansion and
summations of the WKB expansion, order-by-order.   The simplest identification
of such asymptotic expansions is to find many terms explicitly, including
the asymptotic behavior of the coefficients, and test their accuracy
order-by-order with exact results.   Because of the extreme accuracy
of modern asymptotic methods, these comparisons have involved 33 decimal places in similar cases (linear half-well).

To date, only simple analytic forms have been studied: the harmonic oscillator,
particle in a box, the Poschl-Teller well, and the linear half-well \cite{B20b},
all of which have special properties due to their analytic forms.
There are many special cases where quasi-analytic solutions are
known, such as Ref. \cite{X12}, but one needs to be able to 
smoothly approach the semiclassical limit, in which the number of
levels diverges.  Moreover, we seek techniques that ultimately
will be applied to arbitrary (possibly numerically defined) potentials,
so those with analytic solutions might always be special cases. 
The quartic oscillator model studied here contains simple single- and
double-well structures that provide numerous examples of parabolic
minima (the most generic case) that have no analytic solutions, making
them ideal for application of these new methods, but only if extremely
accurate results are easily available.

While it may appear that the results in this paper could be easily 
generated using Mathematica with a single desktop in a short time,
the usefulness of this work is in the careful benchmarking of the
results, the combined analysis of many different aspects, and the
inclusion of asymptotic results, which are unfamiliar to many
computational scientists.  But the greatest value is likely to
be the ability of
the many disparate theorists in many fields to extract highly
accurate results instantly, without having to reperform the
calculations \cite{PM20}.

\sec{Method}

Our Schr\"odinger equation is (in units where $\hbar=m=1$)
\begin{equation}
\label{SHeq}
-\frac{1}{2} \frac{d^2\psi}{dx^2} + v(x) \psi(x) = \eps \psi(x),
\end{equation}
so all energies are in Hartrees, all distances in Bohr radii.
We expand the eigenfunctions in a basis of harmonic oscillator states,
where $\om$ can be freely chosen.  
The
Hamiltonian is pentadiagonal, with only a few non-zero matrix elements
no more than 2 double-steps off the diagonal.  
The nonzero matrix elements of the Hamiltonian
in the harmonic basis are $H_{n,n+2k} = h_k \sqrt{n_{2k}} /16 \om^2$ where 
$h_2 = 1$ and 
\begin{align}
\begin{split}
h_0 =& 4\om (\om^2 - \lam)(2n + 1) + 3(2 n^2 + 2 n + 1),\\
h_1 =& 2 [2 n + 3 - 2\om (\lam + \om^2)],\\
\end{split}
\end{align}
and we use the shorthand
\begin{equation}
\a_p = \prod_{m = 1}^{p}(\a + m), \qquad \a_0 = 1.
\end{equation}

We closely follow Ref. \onlinecite{B19} and use the Eigensystem function in Mathematica
to diagonalize this matrix for various  values of $\lam$
and choices of $\om$ \cite{W20}.  
We denote by $N_B$ the number of basis functions included in
the calculation (both odd and even,
since we did not take advantage of parity).  
Our default choice of $[\om/N_B]$ is [2/200]
but we use [2/400] as a baseline for `exact' energies, and
report errors relative to those values.

A special case is $\smash{\eps = 0}$ for the ground state (magenta
in Fig. \ref{fig:Fdens}).  
This happens at $\smash{\lam = \lam_c}$ which we found
using a golden section search to 
be $\smash{1.3982585455298955302585947187218312604396}$, 
at which the ground state energy is
$-3.955 \times 10^{-41}$.
For a different way of finding energies of oscillators of order $x^{2M}$ using exact quantization conditions see Refs. \cite{V94,V99}; for an approach using lower bounds see Ref. \cite{PM20}.

\sec{Results}

In this section, we report many different results that may be of interest
to different communities under different circumstances.
In each case, we also provide a minimal analysis.

\ssec{Energetics for different potentials}

Here, we simply survey the behavior of the energies and eigenfunctions
for various values of $\lambda$.  Our focus is primarily on positive values
of $\lambda$, which produce the Mexican hat double-well potential.

\begin{table}[htb]
\begin{tabular}{|c|r|r|r|r|}
\hline
n & \multicolumn{1}{c|}{$\lambda  = -1$} & \multicolumn{1}{c|}{$\lambda = 0$} &\multicolumn{1}{c|}{$\lambda  = 2$} & \multicolumn{1}{c|}{$\lambda = 4$} \\
\hline
0 & 0.62092703 & 0.42080497 & -0.29952137 & -2.66144807 \\
1 & 2.02596616 & 1.50790124 & 0.04637108 & -2.65173172 \\
2 & 3.69845032 & 2.95879569 & 1.22797281 & -0.51029304 \\
3 & 5.55757714 & 4.62122032 & 2.45984143 & -0.18078943 \\
4 & 7.56842287 & 6.45350993 & 3.93826197 & 1.16951434 \\
5 & 9.70914788 & 8.42845388 & 5.58129195 & 2.36439189 \\
6 & 11.96454362 & 10.52783077 & 7.36888889 & 3.83579483 \\
7 & 14.32326520 & 12.73833694 & 9.28322263 & 5.44300452 \\
8 & 16.77645279 & 15.04975293 & 11.31134968 & 7.18323497 \\
9 & 19.31695430 & 17.45393416 & 13.44312537 & 9.03984811 \\
\hline
\end{tabular}
\caption{The energies at various values of $\lam$.  See Table S1 for more values of $\lam$, more states, and more digits.}
\label{tab:TE0s}
\end{table}

Our first results are the energetics of the first several eigenstates of
the generalized quartic oscillator.  These values are given to 8 digits
in Table \ref*{tab:TE0s} for four values of $\lam$.  In Table S1 in the
supplementary information, we give 40 digits for 9 values of $\lambda$
for the first 20 eigenvalues.  Here $\lambda=0$ corresponds to the
pure quartic oscillator.  As $\lambda$ grows, the eigenvalues inside the
double well come in pairs, with ever smaller splitting.

\begin{figure}[!htb]
\subfigure{\includegraphics[scale=0.20]{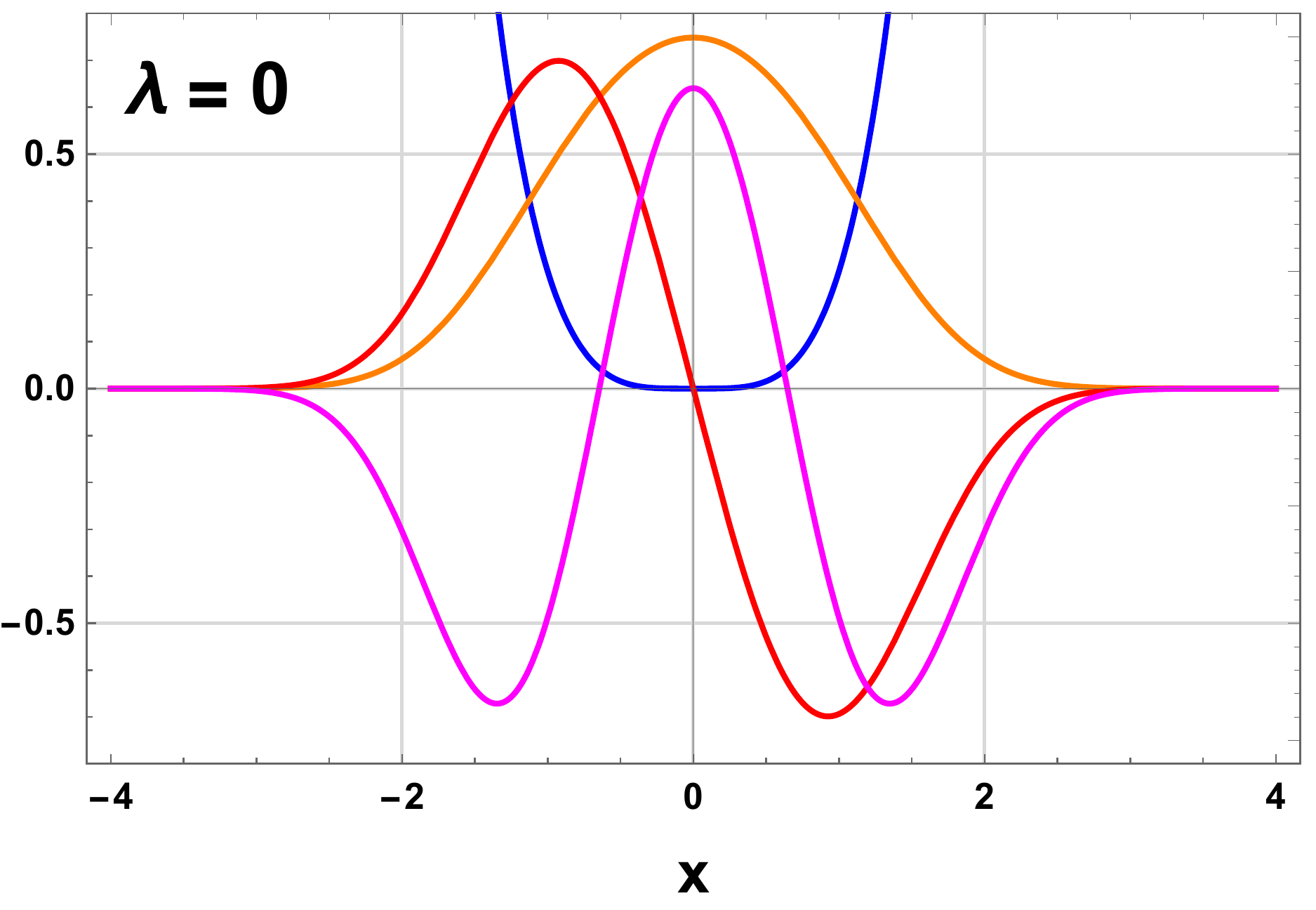}}
\subfigure{\includegraphics[scale=0.20]{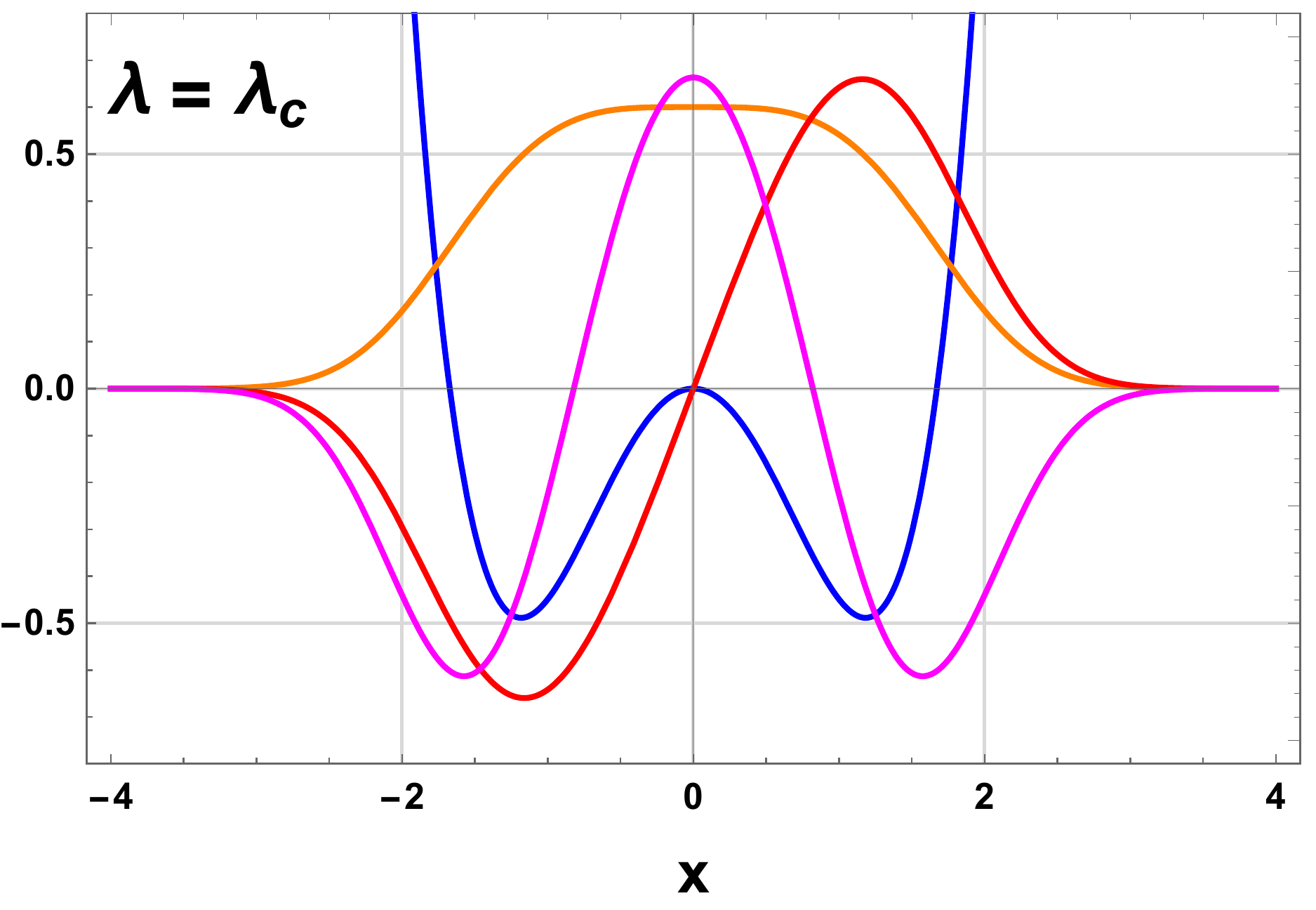}}
\subfigure{\includegraphics[scale=0.20]{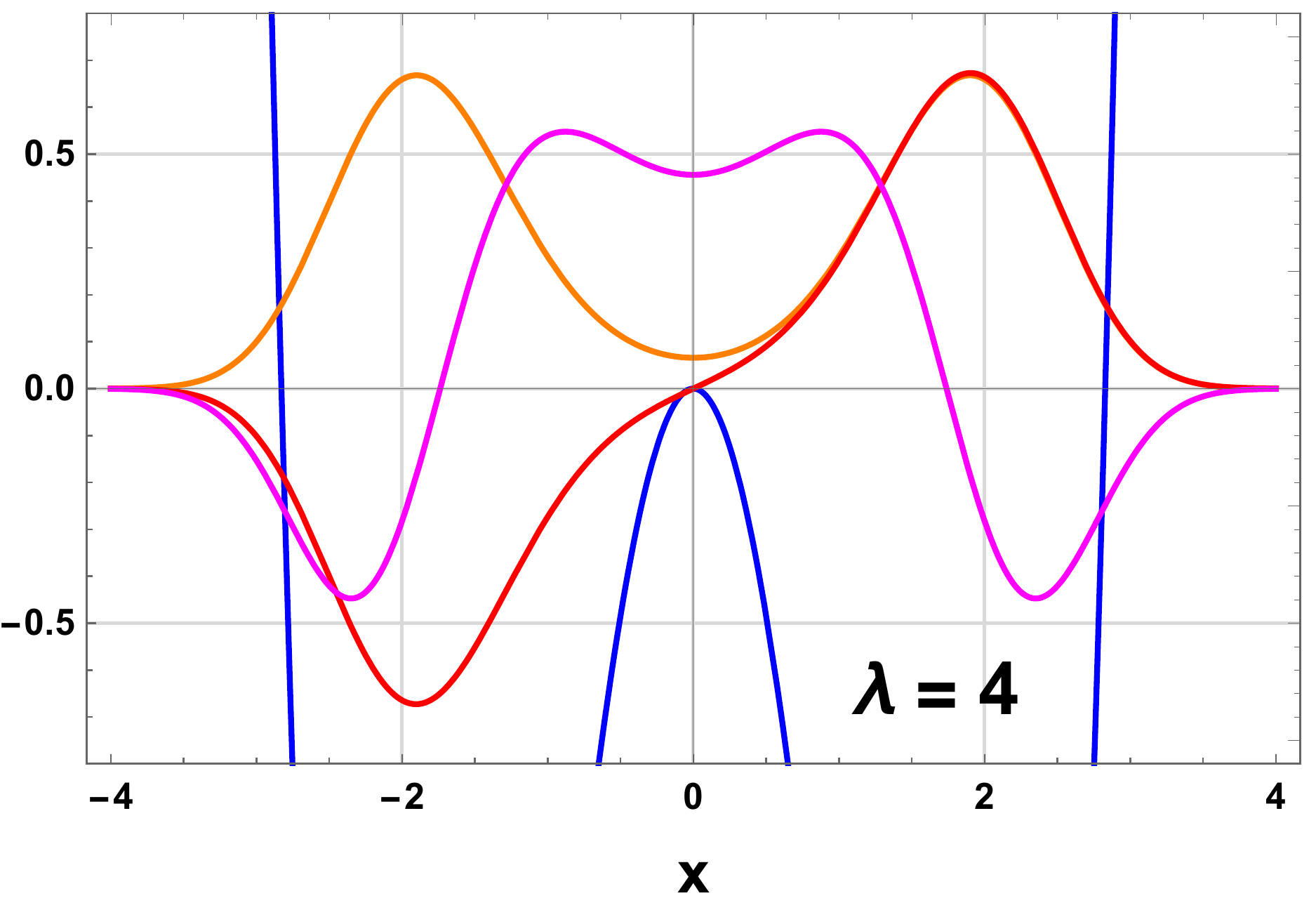}}
\subfigure{\includegraphics[scale=0.20]{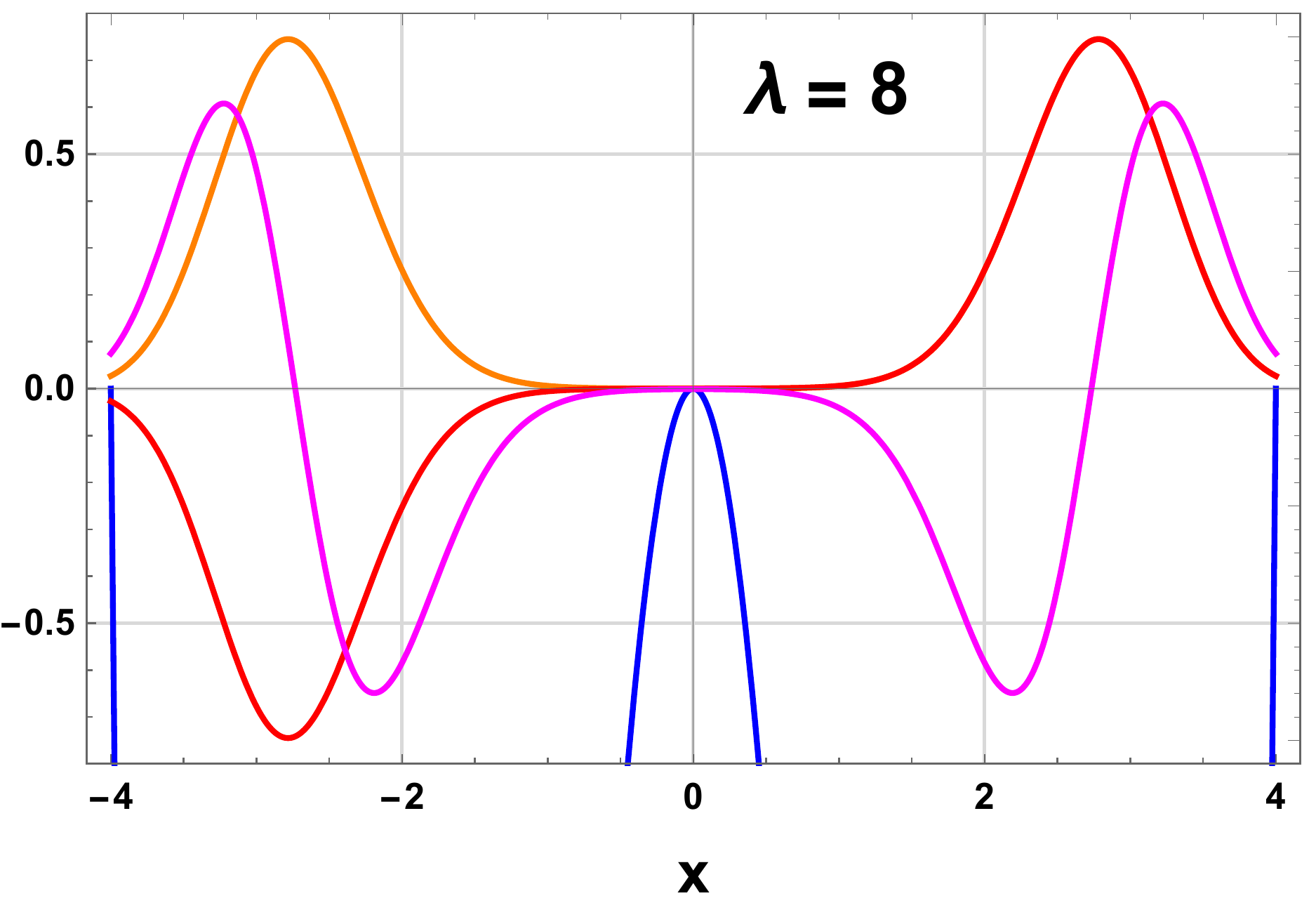}}
\caption{First three eigenfunctions (orange, red, magenta) with potentials (blue) at various values of $\lam$.}
\label{fig:WF}
\end{figure}

We also show the first three stationary states and potentials at
various values of $\lam$ in Fig. \ref{fig:WF}.
As $\lambda$ grows, the
ground-state wavefunction develops a minumum at the origin,
and the first excited state almost matches it in the bulk of the
minimum.   By $\lambda=8$, the wavefunctions are almost indistinguishable,
except for their sign.

\begin{figure}[!ht]
\subfigure{\includegraphics[width=0.8\columnwidth]{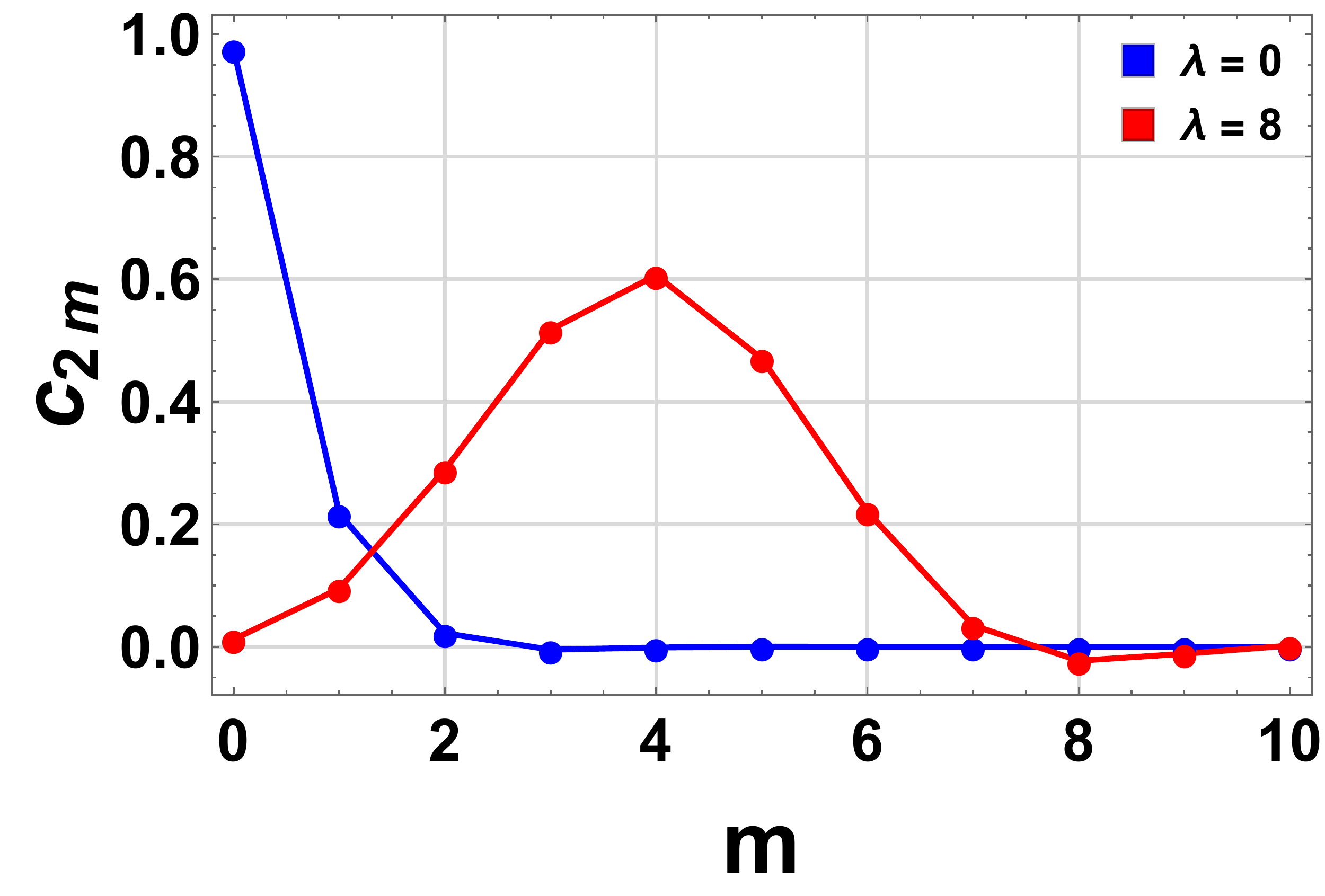}}
\subfigure{\includegraphics[width=0.8\columnwidth]{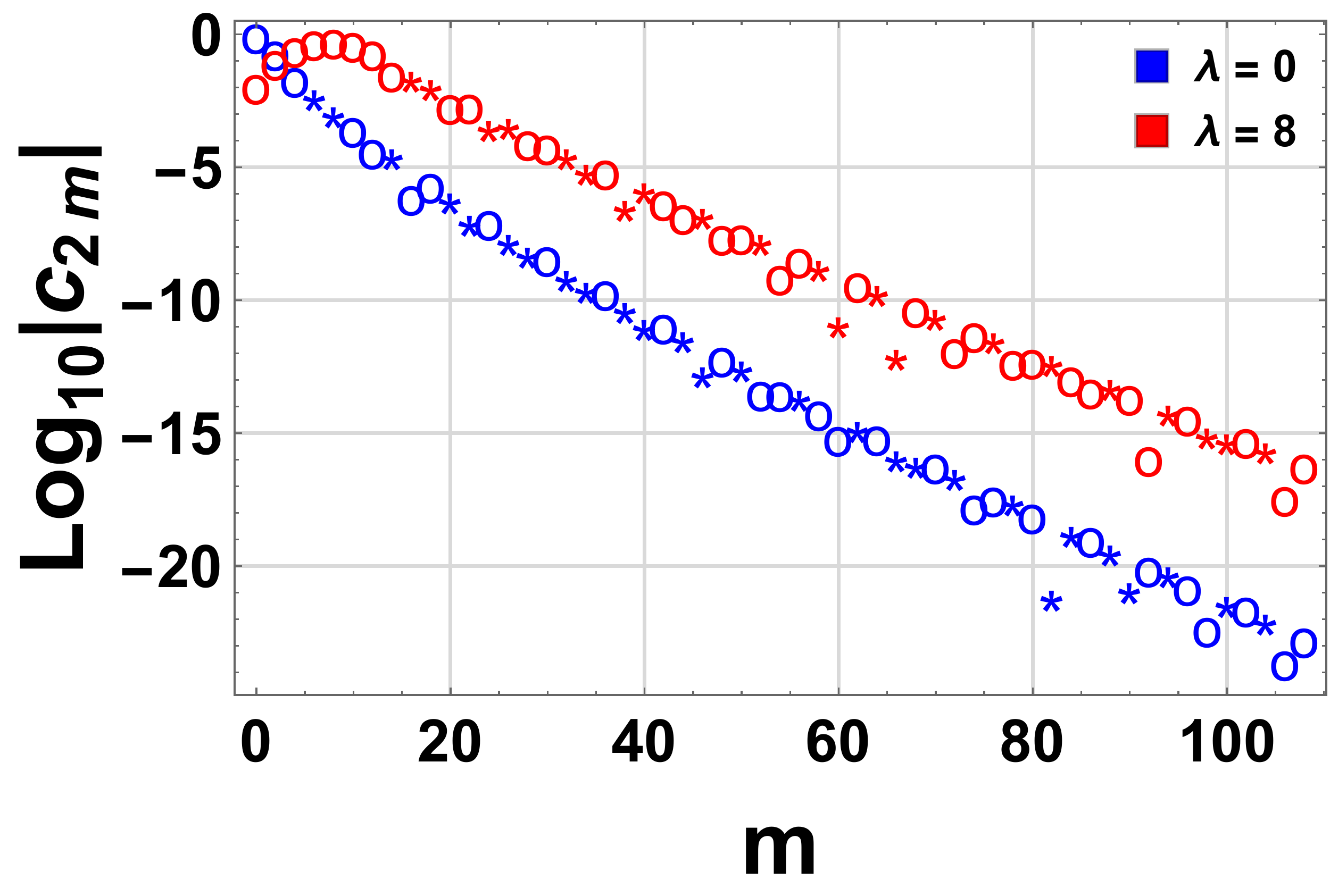}}
\caption{Behavior of the coefficients of the ground-state wave function for the pure quartic
oscillator (blue) and double-well potential (red, $\lambda=8$) in the basis [2/200].  In the lower panel stars and open circles denote $c_{2m}$ of opposite signs.  See Table S2 for more digits.}
\label{fig:WFC}
\end{figure}

In Fig. \ref{fig:WFC} we show the overlap $c_m$
of the ground-state wave function with even oscillator states in
a basis of [2/200] for two values of $\lam$. 
The pure quartic oscillator is dominated by the
ground-state of the harmonic oscillator,
with overlap close to 1,
but the magnitude of the double-well
coefficients grows before ultimately decaying.
In the lower panel, we show that the overlaps decay 
exponentially, but with varying signs.  The broken symmetry
well has overlaps that decay significantly more slowly (about
5 orders of magnitude larger).

\begin{figure}[!ht]
\includegraphics[width=0.8\columnwidth]{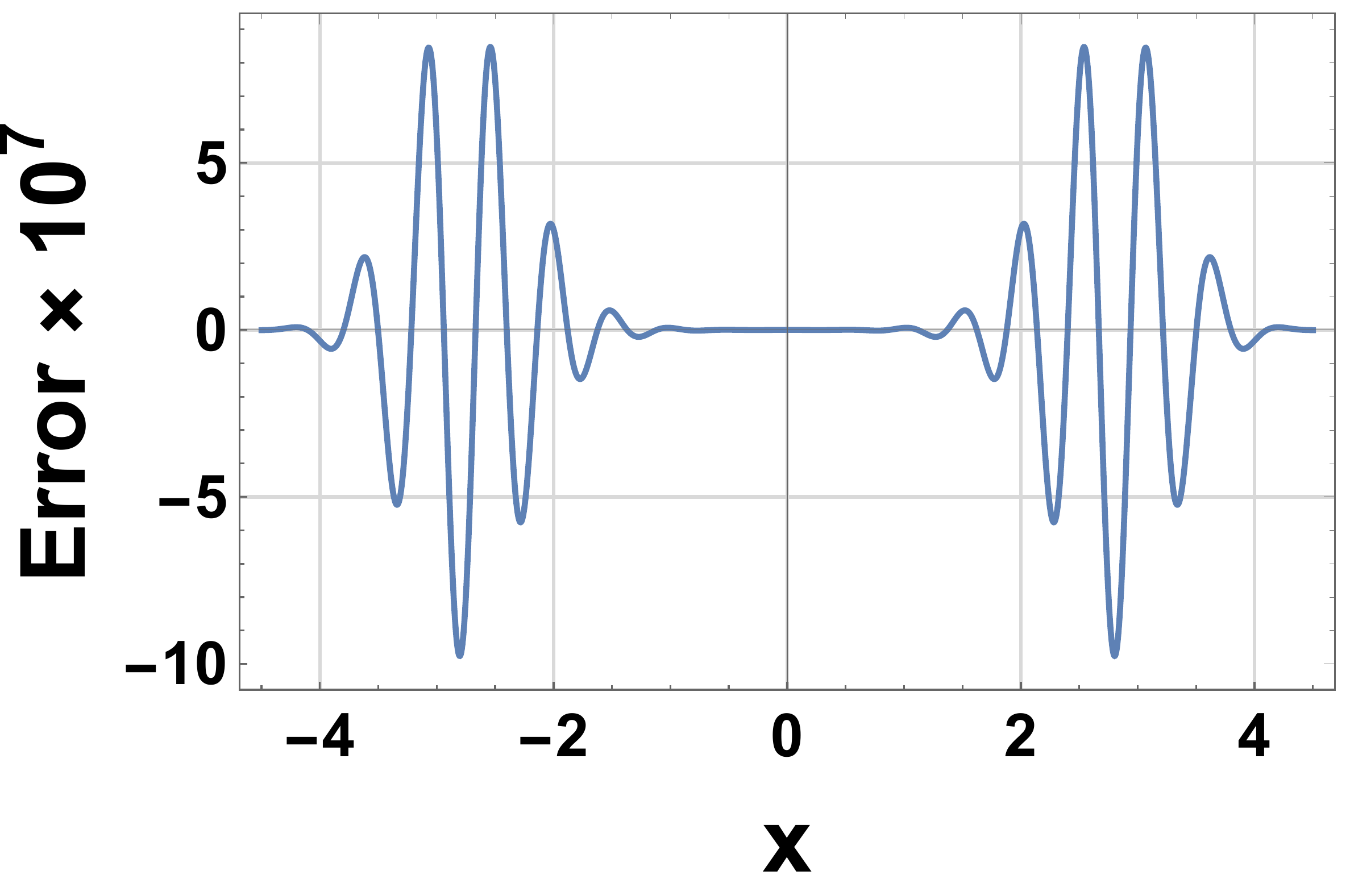}
\caption{The error in the ground state density for $\lam = 8$ calculated with the first 20 coefficients $c_{2m}$ in Table S2.}
\label{fig:ED}
\end{figure}

Lastly, we plot the error
in the ground-state density calculated with the first
20 coefficients of [2/200] in Fig. \ref{fig:ED} for the double
well potential ($\lam = 8$).  
This is not the error of the basis set, but simply the
error caused by truncation after 20 levels.  The error
is very small, oscillates in space, and is localized
in the two different wells.

\ssec{Satisfaction of virial theorem}
\label{virial}

The virial theorem \cite{G05} is a useful check on the accuracy of eigenstates
in a basis.   It is particularly simple here, as the potential
is a sum of two powers of $x$.
For $v_\lam(x)$, the virial theorem requires, for any 
eigensolution
\begin{equation}
\label{vir}
\expval{p^2} + \lambda \expval{x^2}  = \expval{x^4},
\end{equation}
with nonzero matrix elements
\begin{align}
\label{MatrixElem}
\begin{split}
\bar{x}^2_0 =& \bar{p}^2_0 = 2n + 1, \hspace{1mm} \bar{x}^2_1 = -\bar{p}^2_1 = \sqrt{n_2},\\
\bar{x}^4_0 =& 3(2n^2 + 2n + 1), \hspace{1mm} \bar{x}^4_1 = 2 \sqrt{n_2} (2n + 3), \hspace{1mm} \bar{x}^4_2 = \sqrt{n_4},\\
\end{split}
\end{align}
where $\bar{x}_k = x_{n,n+2k} \sqrt{2 \om}$
and $\bar{p}_k = p_{n,n+2k} \sqrt{2/\om}$.  
In particular, at $\lambda_c$, we find the simple formula:
\begin{equation}
\label{vircrit}
\frac{\expval{p^2}}{\expval{x^2}} = \frac{\lam_c}{3}.
\end{equation}
In Table \ref{tab:SVT} we show how closely
our solutions satisfy Eq. \ref{vir}.  
This confirms that with [2/200] we have a very good
approximation to the exact ground states.
Eq. \ref{vircrit} is satisfied to 39 decimal places.

\begin{table}[!ht]
\scalebox{0.9}{
\begin{tabular}{|c|r|r|r|r|}
\hline
$\lam$  & \multicolumn{1}{c|}{$\expval{p^2}$} & \multicolumn{1}{c|}{$\expval{x^2}$} & \multicolumn{1}{c|}{$\expval{x^4}$}& \multicolumn{1}{c|}{$\expval{p^2} + \lam \expval{x^2} - \expval{x^4}$} \\
\hline
 -1 & 0.7096226227 & 0.3548402512 & 0.3547823715 & $-1.0\times 10^{-69}$ \\
0 & 0.5610732993 & 0.4561199557 & 0.5610732993 & $-3.8\times 10^{-68}$ \\
$\frac{1}{2}$ & 0.4859528308 & 0.5399767422 & 0.7559412019 & $-3.2\times 10^{-66}$ \\
1 & 0.4187530838 & 0.6673186910 & 1.0860717748 & $9.6\times 10^{-67}$ \\
$\lambda _c$ & 0.3828873103 & 0.8214946618 & 1.5315492412 & $0.0\times 10^{-40}$ \\
2 & 0.4053838252 & 1.2071184727 & 2.8196207705 & $1.2\times 10^{-63}$ \\
4 & 1.2230281089 & 3.5787191485 & 15.5379047030 & $9.7\times 10^{-60}$ \\
8 & 1.9338080508 & 7.7414002199 & 63.8650098103 & $-1.6\times 10^{-51}$ \\
\hline
\end{tabular}}
\caption{Expectation values and their virial sum for different wells with [2/200].  See Table S3 for more digits.}
\label{tab:SVT}
\end{table}

\ssec{Tunneling between wells}

\begin{figure}[!ht]
\includegraphics[width=0.8\columnwidth]{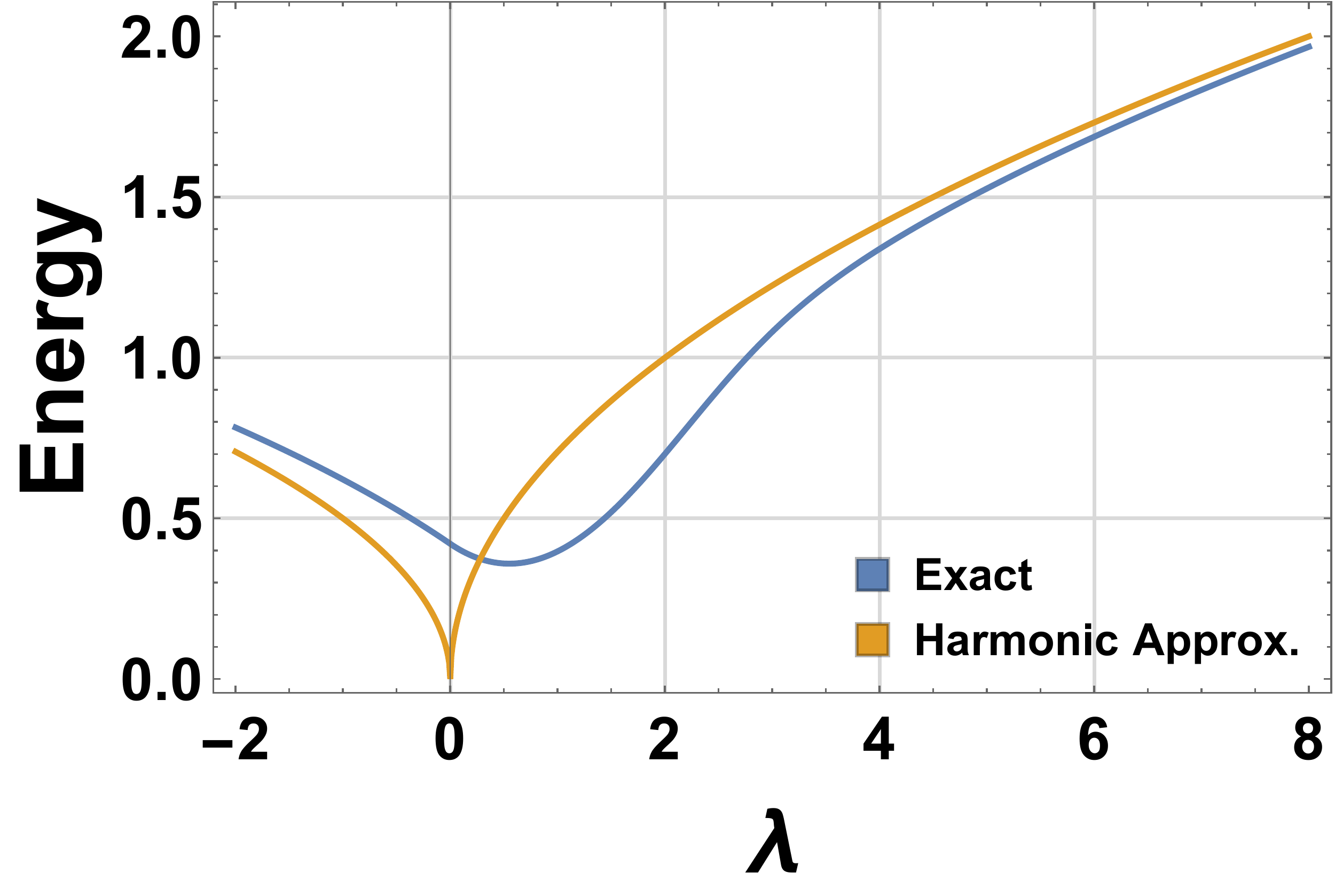}
\caption{Exact zero point energy and it's harmonic approximation.  See Table S4 for many digits.}
\label{fig:ZP}
\end{figure}

\begin{figure}[!ht]
\subfigure{\includegraphics[width=0.8\columnwidth]{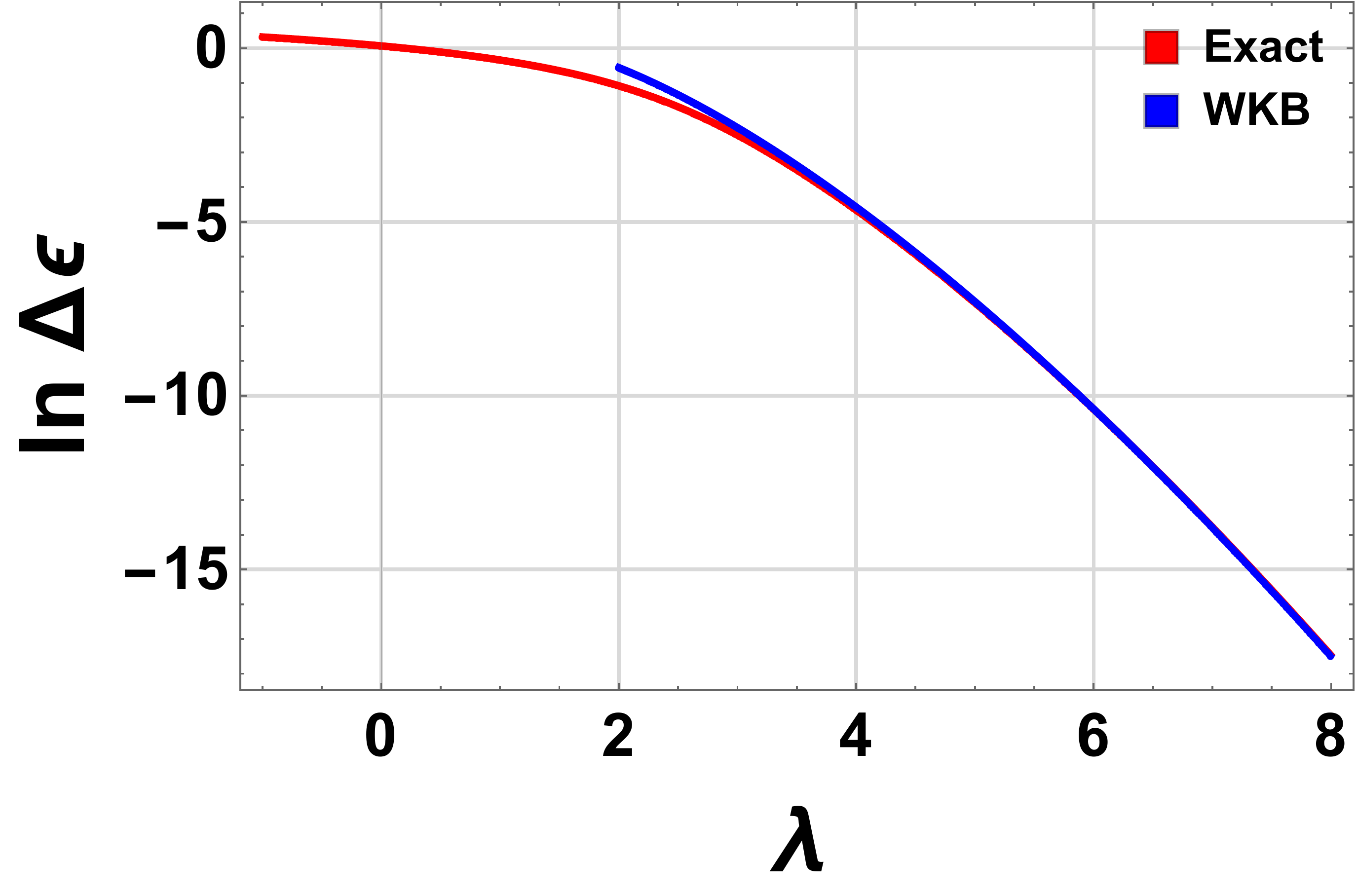}}
\subfigure{\includegraphics[width=0.8\columnwidth]{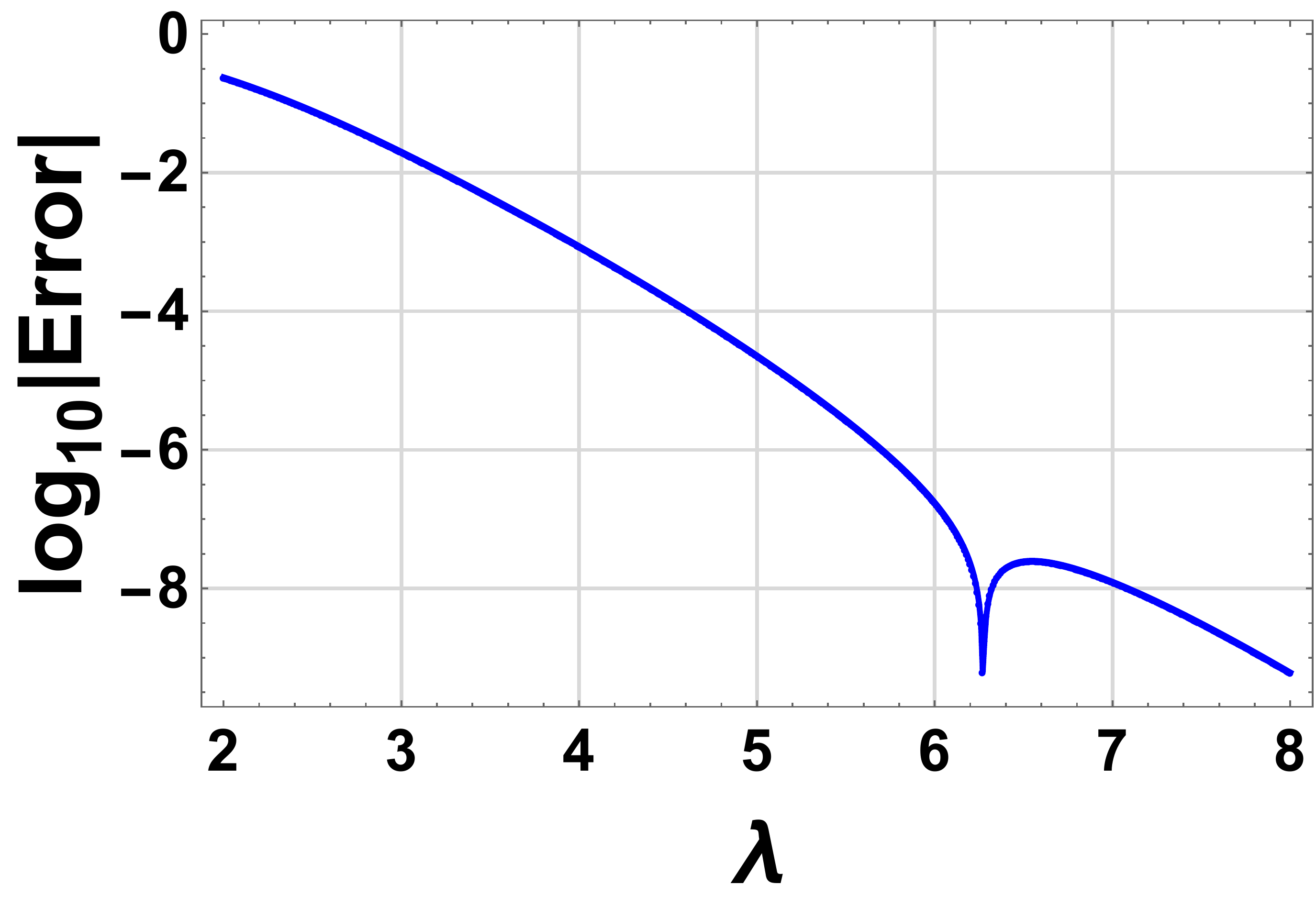}}
\caption{The upper panel compares the exact $\D\eps$ splitting with its WKB approximation in Eq. \ref{AWKB}.  The lower panel shows the error of the WKB approximation.  See Table S4 for the exact $\D\eps$ values.}
\label{fig:E1}
\end{figure}

In this section, we examine both the
zero point energy and the tunneling between the symmetric wells that
occur for positive $\lambda$. 
As mentioned before, the vibrational frequency is $\sqrt{|\lambda|}$
for negative $\lambda$, and ${\sqrt{2\lam}}$ for positive $\lambda$.
Fig. \ref{fig:ZP} shows the exact zero-point energy and it's harmonic
approximation, which becomes accurate as
$|\lambda|$ grows.  

Less trivial is the tunneling between the broken-symmetry wells.
A simple WKB analysis\cite{G05} yields
\begin{equation}
\label{GrifWKB01}
\eps_{\pm} = \frac{\om_0}{2}  \mp \frac{\om_0}{2\pi} e^{-\phi},
\end{equation}
for the lowest two levels, where $\om_0$ is the vibrational
frequency, and $\phi$ is the decay rate for tunneling, evaluated on the ground-state energy.  The splitting is
\begin{equation}\label{13}
\D\eps = \frac{\om_0}{\pi} e^{-\phi(\lam)},
\end{equation}
and $\om_0 = \sqrt{2\lam}$ in the harmonic approximation.  Here $\phi$ is the integral of the absolute value of the momentum $p(x) = \sqrt{2[\eps - v_\lam(x)]}$ over the classically forbidden region between the two wells
\begin{equation}
\label{CFI}
\phi = 2 \int_{0}^{x_1} dx \sqrt{2[v_\lam(x) + \lam^2/4 -\om_0/2]},
\end{equation}
where $x_1=\sqrt{\lambda - 2^{3/4}\lambda^{1/4}}$ is the inner turning point and
$-\lam^2/4 + \om_0/2$ is the harmonic approximation to the ground state energy.
For the approximation to be meaningful, the inner turning point must be
positive, so that $\lambda > 2$.
The appendix shows how to find the asymptotic behavior of the splitting for
large $\lambda$:
\begin{equation}
\label{AWKB}
\D\eps = \frac{2^{11/4} \sqrt{e}}{\pi} \lam^{5/4} \exp(- \frac{(2\lam)^{3/2}}{3}).
\end{equation}
Fig. \ref{fig:E1} shows just how accurate this approximation is.  We have confirmed this expansion numerically.

\ssec{Sextic oscillator}
\begin{table}
$\begin{array}{|c|r|}
\hline
\text{n} & \multicolumn{1}{c|}{\text{Energy}} \\
\hline
0  &  0.43493087870825459239874279292555363392774 \\
1  &  1.64831106336517093605783724089979389227058 \\
2  &  3.44702671416130810318518311192928729955987 \\
3  &  5.67413742993262212079377794412021425830389 \\
4  &  8.24959888596347452014123953299512400416730 \\
5  & 11.13145828009733275940992958109248369395669 \\
6  & 14.28988270823523783646886992792593890944806 \\
7  & 17.70235221954562079900780369145795203975459 \\
8  & 21.35111714819949016424850927302262003053565 \\
9  & 25.22171285703672981248414385975444098122878 \\
10 & 29.30205319182515264341685743177538338413530 \\
11 & 33.58184072441714447659452964789566752711547 \\
12 & 38.05216382472115780004920306874727674910548 \\
13 & 42.70521061321923679887306117322169371214762 \\
14 & 47.53405945114494426110319556943507798525146 \\
15 & 52.53252145969695924991153932257382502470230 \\
16 & 57.69501952928699913809886057200837269497456 \\
17 & 63.01649360693670075261996021857703983520713 \\
18 & 68.49232534279718129547616418400704809075654 \\
19 & 74.11827728288342368118468014333735452468298 \\
\hline
\end{array}$
\caption{First twenty energies of the sextic oscillator calculated with [2/800].  The energies are accurate to all 41 digits shown.}
\label{tab:SOE}
\end{table}

In this section, we apply exactly the same technology to finding the energies
of the sextic oscillator:
\begin{equation}
v(x) = \frac{x^6}{6}.
\end{equation}
The quartic and sextic oscillators both belong to the class of potentials whose
exact solutions are given by Heun's special function \cite{LI19}.
In our harmonic basis, the nonzero Hamiltonian matrix elements are
$H_{n,n+2k} = \sqrt{n_{2k}} h^{(6)}_k / 48 \om^3$
where $h^{(6)}_3 = 1$ and
\begin{align}
\begin{split}
h^{(6)}_0 =& (2n + 1) [10 n (n+1)+3 \left(4 \omega ^4+5\right)],\\
h^{(6)}_1 =& 3 [5 n (n+3)-4 \omega ^4+15],\\
h^{(6)}_2 =& 3 (2 n+5),\\
\end{split}
\end{align}
i.e., they go one more step away from the diagonal.
The energies of the first twenty sextic oscillator states are given in Table \ref{tab:SOE}.

\begin{figure}[!ht]
\includegraphics[scale=0.30]{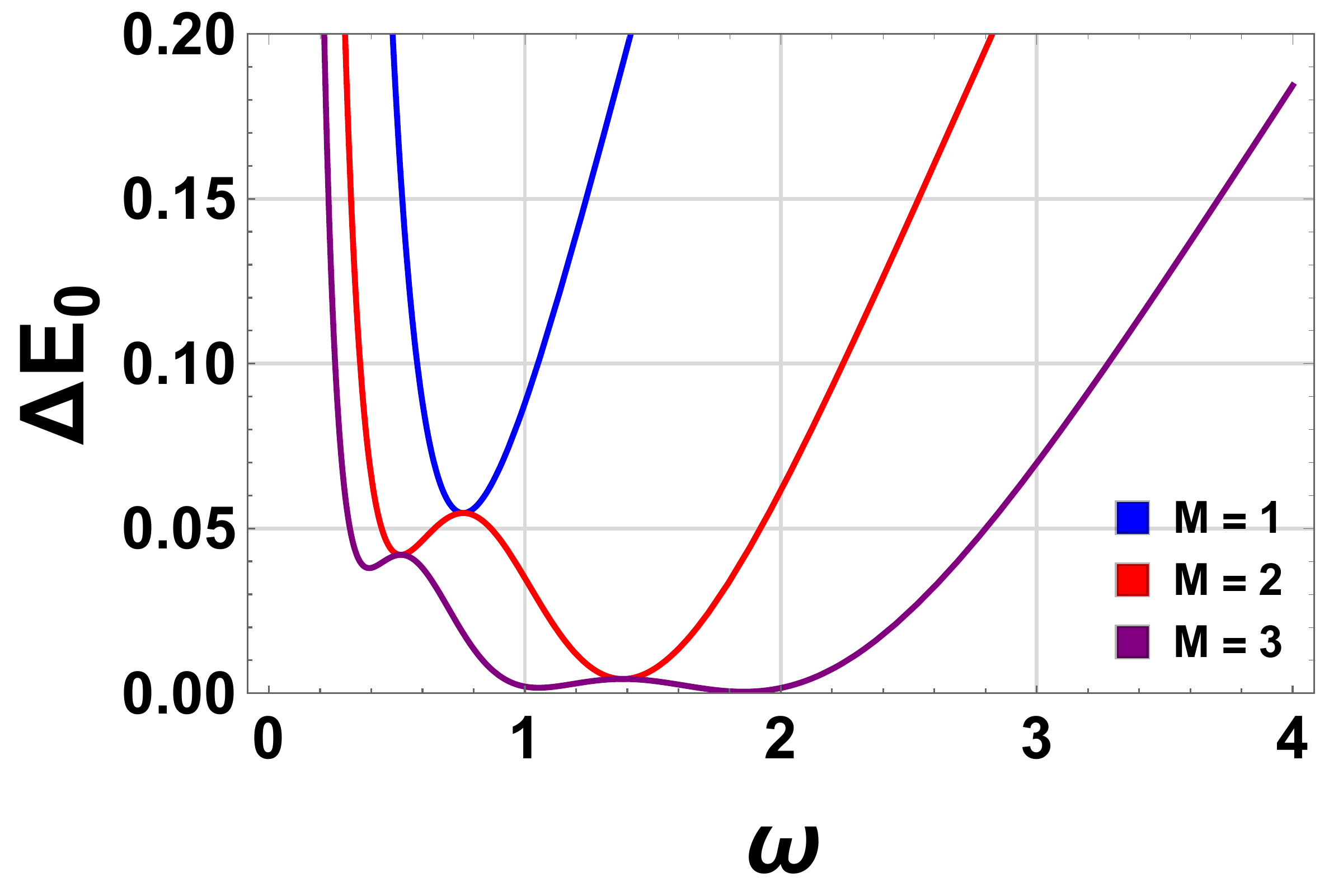}
\caption{The errors of the analytic expressions for the approximate ground state with 1, 2, and 3 even basis functions and $\lam = \lam_c$.} 
\label{fig:ALBE}
\end{figure}

\ssec{Analytic results for a few states}
It can often be useful to find an approximate solution using just a few basis functions,
instead of hundreds.  Here we give analytic formulas for the lowest lying even energies as functions
of $\om$ and $\lam$
when only 1, 2, and 3 even oscillator states are used. 
These expressions can be useful for quick estimates of low-lying eigenvalues.
The approximate ground-state energy with one even basis function is
\begin{equation}
\label{GSB1}
\eps_0=\frac{3}{16 \omega ^2}+\frac{\omega }{4} - \frac{\lambda }{4 \omega }, \qquad (N_B = 1).
\end{equation}
The approximate ground- and second-excited states
with two even basis functions are:
\begin{align}
\label{GSB2}
\begin{split}
\eps_\pm &= \frac{3 \left(\omega ^2-\lambda \right)}{4 \omega } 
+ \frac{21 \pm 2 {\sqrt{D}}}{16 \om^2}, \qquad (N_B = 3),\\
D &= 8 \omega  [3 \omega  \left(\lambda ^2+ \omega ^4+2 \omega \right)-2 \lambda  \left(\omega ^3+6\right)]+99.\\
\end{split}
\end{align}
With three even basis functions the first three approximate even state energies are ($n=0,2,4$):
\begin{align}
\label{GSB3}
\begin{split}
\eps_n  &= \frac{1}{48 \om^2} \bigg[ 15 (11 - 4 \lam\om + 4 \om^3)\\
        & - (-1)^{\d_{n,2}} 8 \sqrt{6 D} \cos \left(  \frac{\phi}{3} + \frac{(n + 1)\pi}{6} \right)  \bigg], \hspace{1mm} (N_B = 5),\\
D       &= 15 [\omega ^2 \left(\lambda ^2+\omega ^4+\omega \right)-7 \lambda  \omega +13]-2 \lambda  \omega ^4,\\
\sin\phi& = \frac{9 B}{8 \sqrt{6 D} D},\\
B       &= 20 \lambda  \omega  \left(\lambda  \omega  [51-4 \omega  (\lambda +\omega ^2)]+2 [2 \omega ^6+7 (\omega ^3-15)]\right)\\
        &+ 4\omega ^6 \left(20 \omega ^3-57\right) + 5575.\\
\end{split}
\end{align}
At $\lam_c$ (Fig. \ref{fig:ALBE}), the least error in the ground state energy is $5.467 \times 10^{-2}$  at $\om = 0.7595$ with Eq. \ref{GSB1}, $4.320 \times 10^{-3}$ at $\om = 1.383$ with Eq. \ref{GSB2}, and $4.563 \times 10^{-4}$ at $\om = 1.854$ with Eq. \ref{GSB3}.

\begin{figure}[!ht]
\subfigure{\includegraphics[width=0.8\columnwidth]{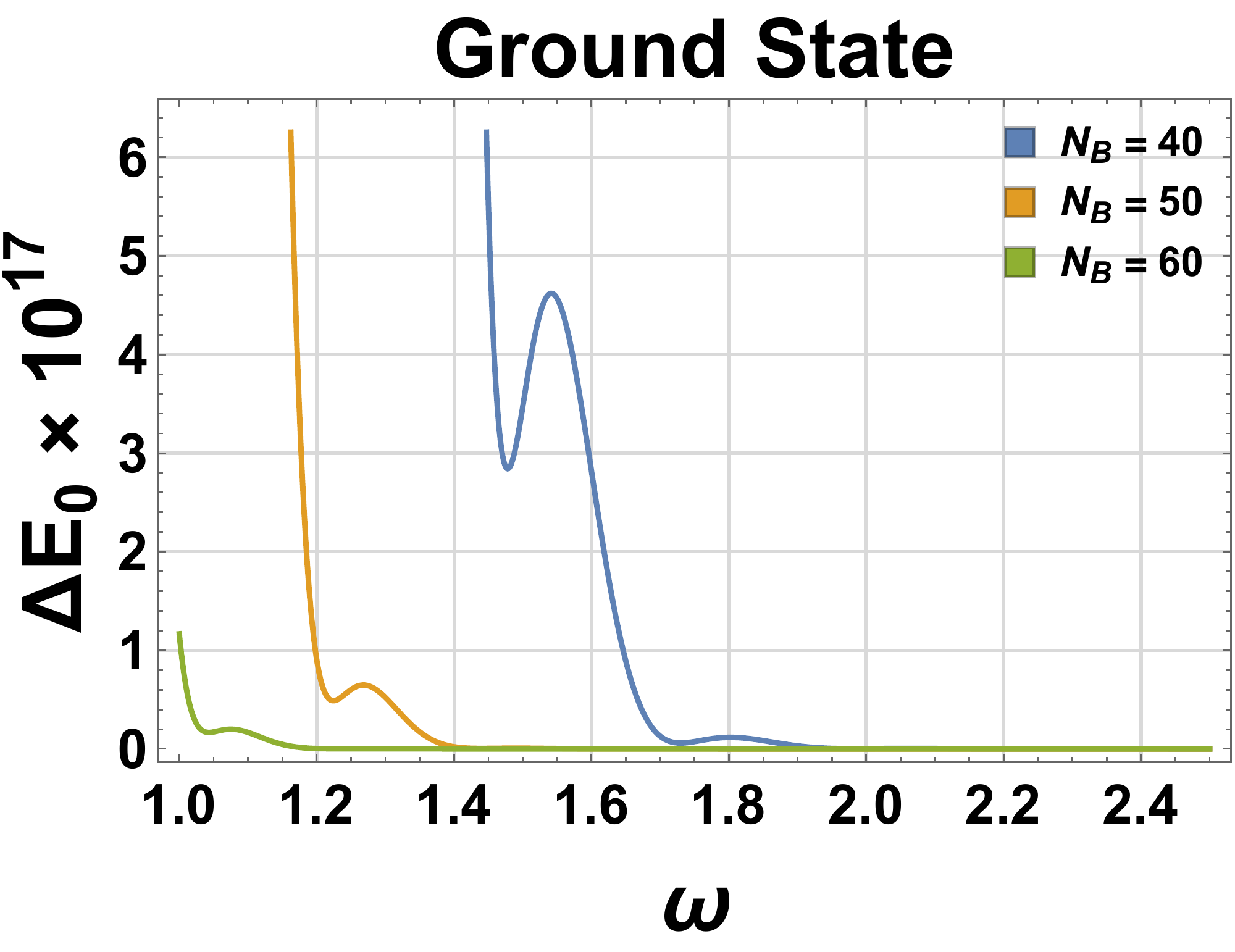}}
\subfigure{\includegraphics[width=0.8\columnwidth]{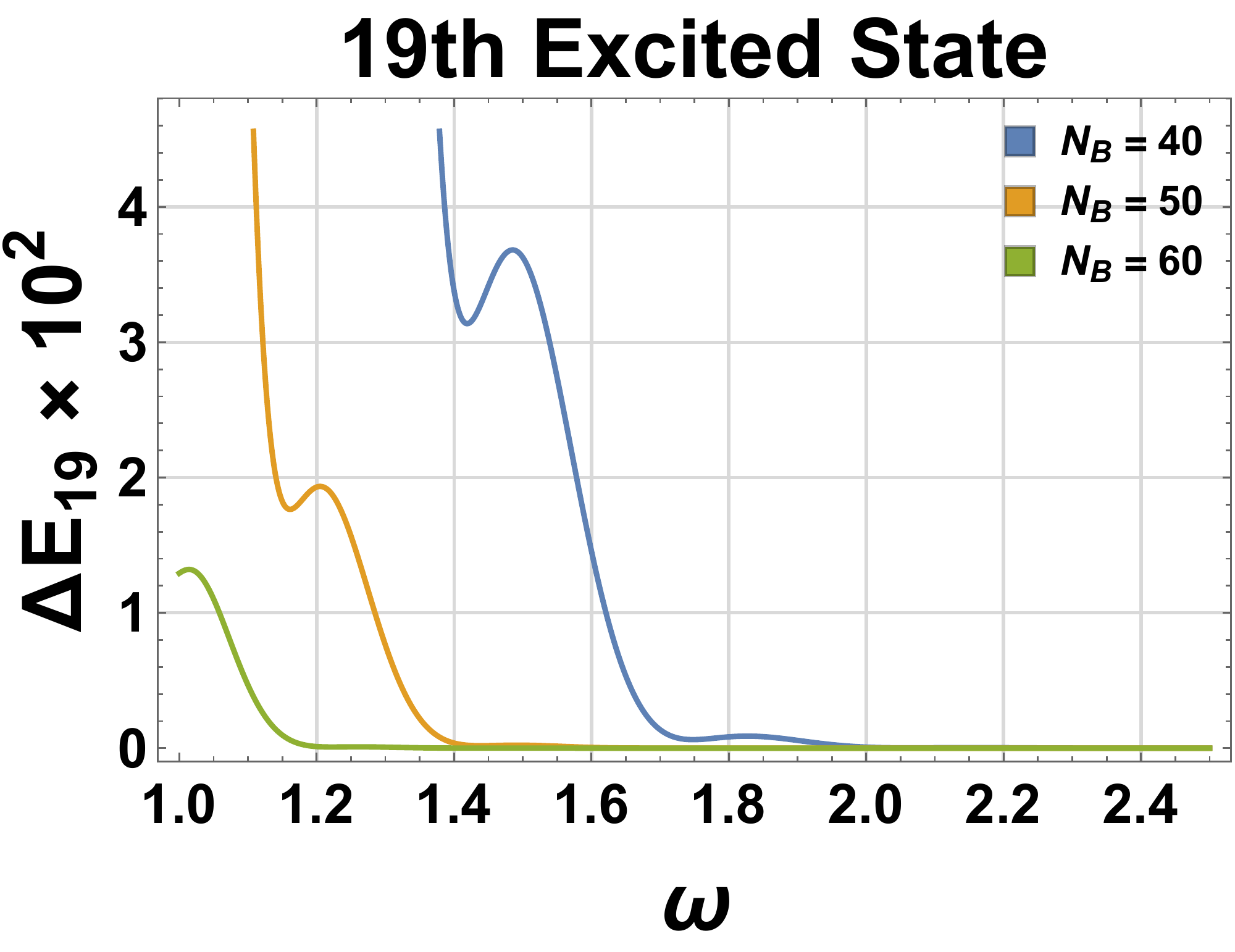}}
\caption{The errors of the ground state and 19th excited state (i.e. 10th odd state) as a function of $\om$.  See Table S5 for more digits.}
\label{fig:VO}
\end{figure}

\ssec{Error dependence on $\omega$}
In this paper we have usually set the basis set angular frequency $\om$ to 2.
Now we analyze what happens to the error of the ground and
a highly excited state of the pure quartic oscillator as $\om$ is varied.
The error as a function of $\om$ for a fixed number of basis states
is complicated and has several local minima, as we found in the
previous section.  Nevertheless there is a clear trend
for the pure quartic oscillator as seen in Fig. \ref{fig:VO}: the error
tends to level off to a very low value as $\om$ increases, though it must increase
if $\om$ becomes too large.  
The error for the 19th excited state is orders of magnitude greater
than that of the ground state with e.g., $N_B=40$.
For our purposes, the value of $\om=2$ yielded sufficient accuracy for the 
basis sets we could afford.

At each minimum as a function of $\omega$, the virial
theorem is exactly satisfied. We could have performed such a search for any
of our calculations.  But as we could achieve sufficient accuracy with fixed
basis sets, we chose the simpler and less computationally intensive route. 
This preserves any cancellation of errors in energy differences, and retains
the virial as a test of accuracy, as in Sec. \ref{virial}.

\begin{figure}[!ht]
\includegraphics[width=0.8\columnwidth]{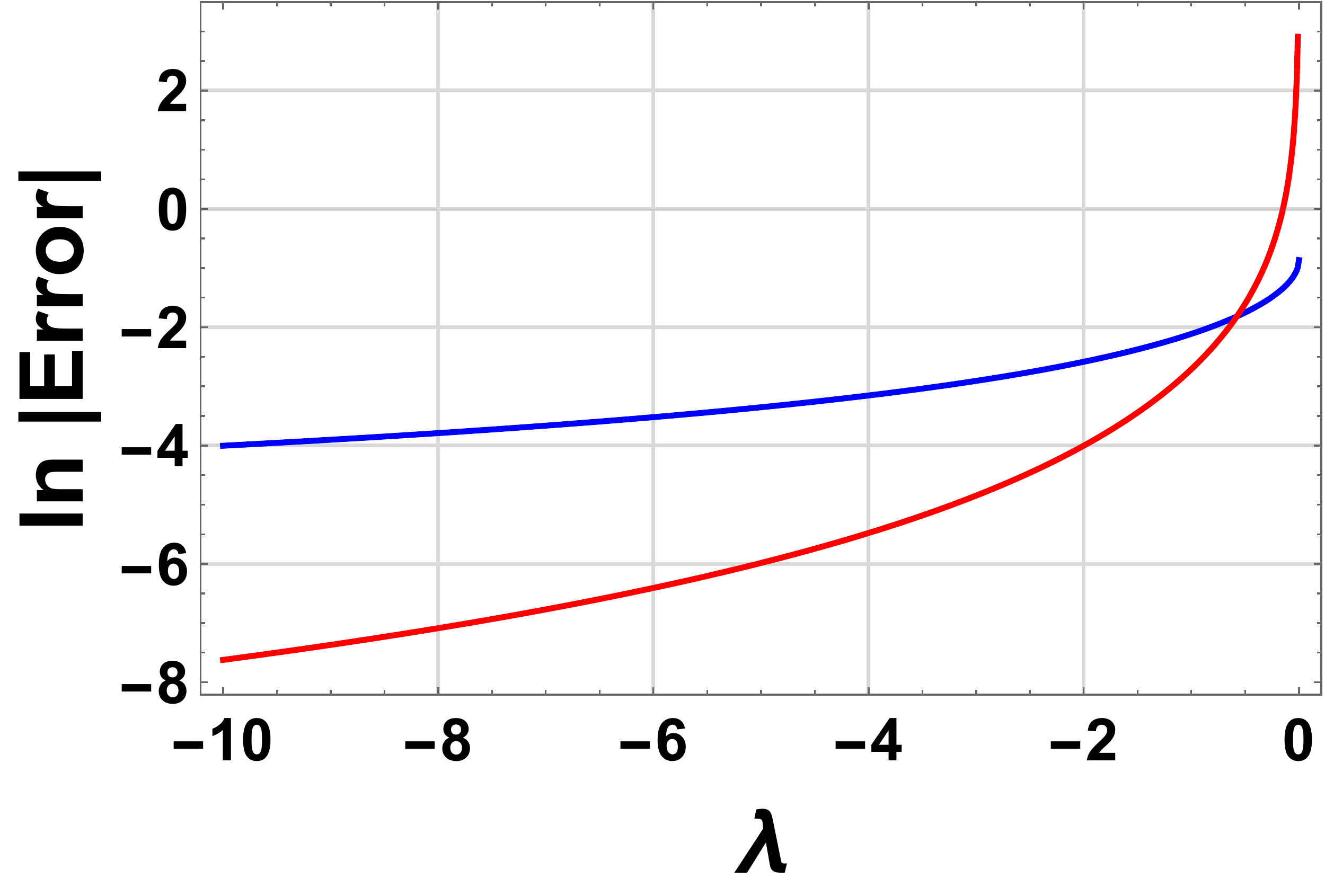}
\caption{The error in the ground state energy from zeroth (blue) and first order (red) perturbation theory.  See Table S6 for accurate numbers.}
\label{fig:PT}
\end{figure}

\ssec{Quartic potential as perturbation}
Consider the case where $\lambda$ is large and negative, and treat the
quartic potential as a perturbation.  This problem and it's analytic structure was studied in Refs. \cite{BW68,BW69,SD70}.  The zeroth, first, and second order contributions to the energies are
\begin{align}
\begin{split}
\eps_n^{(0)} =& \left( n + \frac{1}{2} \right) \sqrt{|\lam|},\\
\eps_n^{(1)} =& \frac{3(2n^2 + 2n + 1)}{16 |\lam|},\\
\eps_n^{(2)} =& - \frac{(1+2n)[17n(n + 1) + 21]}{128 |\lam|^{5/2}}.\\
\end{split}
\end{align}
Fig. \ref{fig:PT} shows the resulting error in the ground state energy.

\ssec{Asymmetric wells}
\begin{table}
\scalebox{0.8}{
$\begin{array}{|c|r|r|}
\hline
\text{n} & \multicolumn{1}{c|}{\text{Energy}}  & \multicolumn{1}{c|}{\text{difference}} \\
\hline
0  & -2.84178633947585932025083089644391799927051 & -0.1803382726 \\
1  & -2.47315425631802332765380889849256442811375 &  0.1785774606 \\
2  & -0.55873199537080776530382207755703245716816 & -0.0484389572 \\
3  & -0.13874405574419168918357404272418272068978 &  0.0420453773 \\
4  &  1.16447030692387601517830930663899021569773 & -0.0050440297 \\
5  &  2.36573532391604302707250344355822369758964 &  0.0013434316 \\
6  &  3.83568104437914171468097998090919165511556 & -0.0001137882 \\
7  &  5.44302728549612071643675096706241920608503 &  0.0000227623 \\
8  &  7.18320046060050124539642226054074137331752 & -0.0000345135 \\
9  &  9.03979350957144560421088106446755353161811 & -0.0000546026 \\
10 & 11.00244857292039554353678211539683848606553 & -0.0000702608 \\
11 & 13.06271608472508976557198671004627390279753 & -0.0000808253 \\
12 & 15.21369451941728871017509326975221513855338 & -0.0000881381 \\
13 & 17.44958477274837116578655941806092145442188 & -0.0000931621 \\
14 & 19.76542796609356112334994801927133021018867 & -0.0000965616 \\
15 & 22.15692272774436386892142052150687339550426 & -0.0000987886 \\
16 & 24.62029451810540168163027589352125330801114 & -0.0001001588 \\
17 & 27.15219883683433104396784761809263032892086 & -0.0001008965 \\
18 & 29.74964765111306868111782309284631425560723 & -0.0001011638 \\
19 & 32.40995226660074725753866259978103913402690 & -0.0001010795 \\
\hline
\end{array}$}
\caption{The first twenty energies when $\lam = 4$ and $\a = 0.1$ calculated with [3/200].  The difference from $\a=0$ is reported.  All energies are accurate to the 41 digits given.}
\label{tab:LPT}
\end{table}

\begin{figure}[!htb]
\includegraphics[width=0.8\columnwidth]{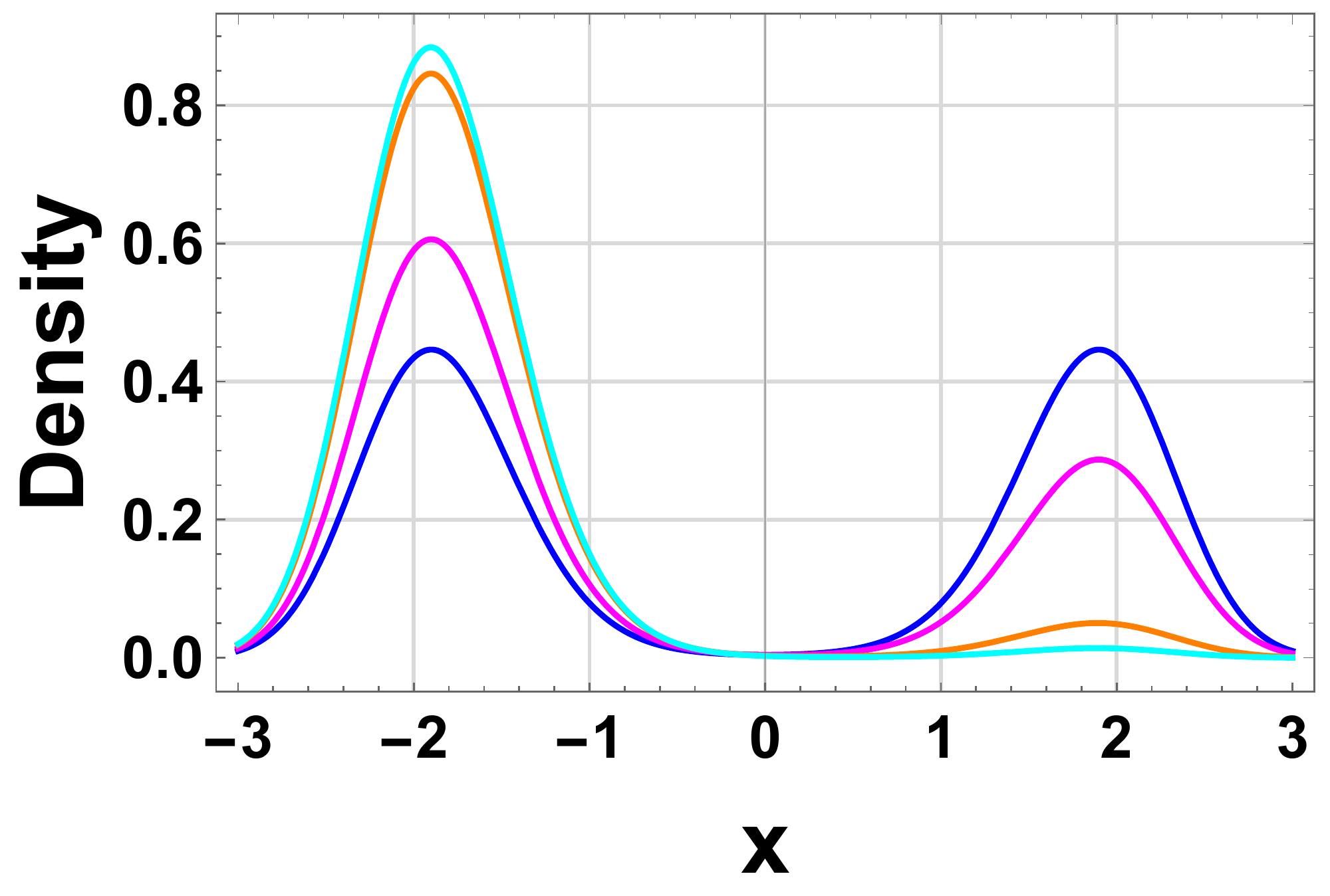}
\caption{The ground state density with $\lam = 4$ and various values of $\a$: 0 (blue), 0.001 (magenta), 0.005 (orange), 0.01 (cyan).}
\label{fig:LPTD}
\end{figure}

We now examine the effect of breaking the symmetry of $v_\lam(x)$ by adding a linear term
\begin{equation}
v_{\lam,\a}(x) = \frac{x^4}{4} - \lam \frac{x^2}{2} + \a x.
\end{equation}
We only examine the case $\lam = 4$.  In Table \ref{tab:LPT}, we show both
the energies for the case $\alpha = 0.1$ and their difference 
from the unperturbed case $\a = 0$.  
As one side of the well is depressed and the other elevated, for the low-lying
states, the differences alternate in sign.  As one goes further up the well, eventually
all states are lower than their symmetric counterparts.

In Fig. \ref{fig:LPTD} we show how the ground state density varies
as $\a$ is increased.  Even a very
small value of $\a$ causes substantial asymmetry in the ground-state
density, with almost all the weight in the lower well when $\alpha=0.1$.

\begin{table}[!ht]
\scalebox{0.8}{
\begin{tabular}{|c|c|c|}
\hline
$n$ & $k_n$ & $l_n$ \\
\hline
0 &  1 & 3 \\
1 & -1 & 4 \\
2 & 11 & $3 \times 2^9$ \\
3 & $7 \times 11 \times 61$ & $3 \times 5 \times 2^{11}$ \\
4 & $- 5 \times 13 \times 17 \times 353$ & $7 \times 2^{19}$ \\
5 & $- 11^2 \times 19 \times 23 \times 1009$ & $3 \times 2^{21}$ \\
6 & $ 5 \times 17 \times 29 \times 49707277$ & $3 \times 11 \times 2^{28}$\\
7 & $3^4 \times 7 \times 19 \times 23 \times 31^2 \times 109 \times 1429$ & $13 \times 2^{30}$ \\
8 & $-7 \times 11 \times 29 \times 37 \times 41 \times 4477909193$ & $3 \times 2^{39}$ \\
9 & $-5 \times 11 \times 19 \times 23 \times 31 \times 43 \times 47 \times 1489 \times 6397 \times 8263$ & $17 \times 2^{41}$ \\
10& $7 \times 29 \times 37 \times 41 \times 53 \times 59 \times 3618497 \times 83558311$ & $3 \times 19 \times 2^{48}$ \\        
\hline
\end{tabular}}
\caption{The known $A_{2n}$ are given by the $k_n$ and $l_n$ in $A_{2n} = (k_n/l_n) \sqrt{\pi} R^{(-1)^n}$ where $R = \G(1/4)/\G(3/4)$.}
\label{A2nWKB}
\end{table}

\begin{table}
\scalebox{0.8}{
$\begin{array}{|c|r|r|}
\hline
\text{n} & \multicolumn{1}{|c|}{A_{2n}} & \multicolumn{1}{|c|}{B_{2n}} \\
\hline
0 &        1.74803836952807987364 &   1.09253465015618881971 \\
1 &       -0.14976752934194902593 &   0.03864050890478489138 \\
2 &        0.03755551184532984104 &  -0.00385400310372957406 \\
3 &        0.09159610485926231443 &  -0.00192912270049430287 \\
4 &       -0.55736698690956972061 &   0.00455965708128336316 \\
5 &       -5.08024277232829207859 &   0.01110530044025892883 \\
6 &       72.53628245858812264379 &  -0.05611473947869961072 \\
7 &     1591.77267377039864942443 &  -0.34434940280222393316 \\
8 &   -48231.49420089254973409982 &   3.51923362542214395736 \\
9 & -1899239.99920378994311897265 &  39.91240769625859000539 \\
10& 95166684.23238064054845710849 &-660.60215595012034723938 \\
\hline
\end{array}$}
\caption{The known WKB coefficients for the pure quartic oscillator reported to twenty digits.}
\label{tab:QWKB}
\end{table}

\begin{table*}
\scalebox{0.6}{
$\begin{array}{|c|c|c|c|c|c|c|c|}
\hline
n & C_n & a_{n,0} & a_{n,1} & a_{n,2} & a_{n,3} & a_{n,4} & a_{n,5}\\
\hline
0 & 9 & 1 &  &  & & &  \\
1 & 1 & 1  &  &  & & & \\
2 & 1/72 & 5 & 11/192 &  & & & \\
3 & 11/972 & -1 & 93/640 &  & & & \\
4 & 17/559872 & -77 & 539/20 & 102829/86016 & & & \\
5 & 23/5038848 & 119 & -3289/48 & 28171999/430080 & & &  \\
6 & 29/1088391168 & 5083 & -661089/160 & 6734014687/716800 & 49829732957/90832896 & &\\
7 & 1/1224440064 & -43355 & 2931929/64 & -10264192781/61440 & 492349052125069/1349517312 & &\\
8 & 41/176319369216 & -164749/4 & 806113/15 & -262775969173/983040 & 787570022698313/527155200 & 45866361756966241/355140108288 &\\
9 & 47/2115832430592 & 3230513/27 & -7446461/40 & 4267944409223/3686400 & -1335041940357576377/120766464000 & 4907566420869344641093/98107454914560 &\\
10 & 53/1371059415023616 & 58397735/3 & -9015402055/256 & 89325797863511/344064 & -955865010579864937/268369920 & 1407560427696573497146789/32702484971520 & 5620192339921634510441141/1187588522115072\\
\hline
\end{array}$}
\caption{The constants yielding the $\beta_n$ via Eq. \ref{BetaPoly}.}
\label{tab:EWKBC}
\end{table*}

\ssec{Asymptotic analysis of pure quartic oscillator}
The asymptotic solution of the pure and generalized quartic oscillator has
been studied many times before \cite{BO99,V83,DP97,BW72}.  We analyze only
the pure quartic oscillator and closely follow Bender \& Orszag \cite{BO99}.
The WKB series for a pure quartic oscillator with potential $v(x) = x^4/4$ yields
the implicit formula
\begin{equation}
\sum_{m = 0}^{\infty} A_{2m} (4 \eps^{3/2})^{1/2-m} = \left(n + \frac{1}{2}\right) \pi,
\end{equation}
with the known $A_{2n}$ reported in Table \ref{A2nWKB} and in Refs. \cite{BO99,V80}.  One can invert this implicit expression to an explicit
formula for each level: 
\begin{equation}
\label{WKBexpl}
\eps_n = 2^{-1/3} \sum_{m = 0}^{\infty} B_{2m}
\left(n + \frac{1}{2}\right)^{4/3 - 2m}.
\end{equation}
We give the known $A_{2n}$ and $B_{2n}$ coefficients numerically in Table \ref{tab:QWKB} to twenty decimal places.  The analytic forms of the $B_{2n}$ coefficients are given by
\begin{equation}
B_{2n} = (-1)^{\floor{n/2}} \frac{\pi^{2-n} \beta_{n}}{18^{1/3} \G(1/4)^{8/3}},
\end{equation}
where the $\beta_n$ are polynomials of order $\floor{n/2}$ in $\g$:
\begin{equation}
\label{BetaPoly}
\beta_n = C_n \sum_{k = 0}^{\floor{n/2}} a_{n,k} \g^k,
\end{equation}
where $\smash{\g = \Gamma(1/4)^8/\pi^4}$.  This
allows the 11 known $\beta_{2n}$ to be given by the constants in Table \ref{tab:EWKBC}.

\begin{figure}[!ht]
\includegraphics[width=0.8\columnwidth]{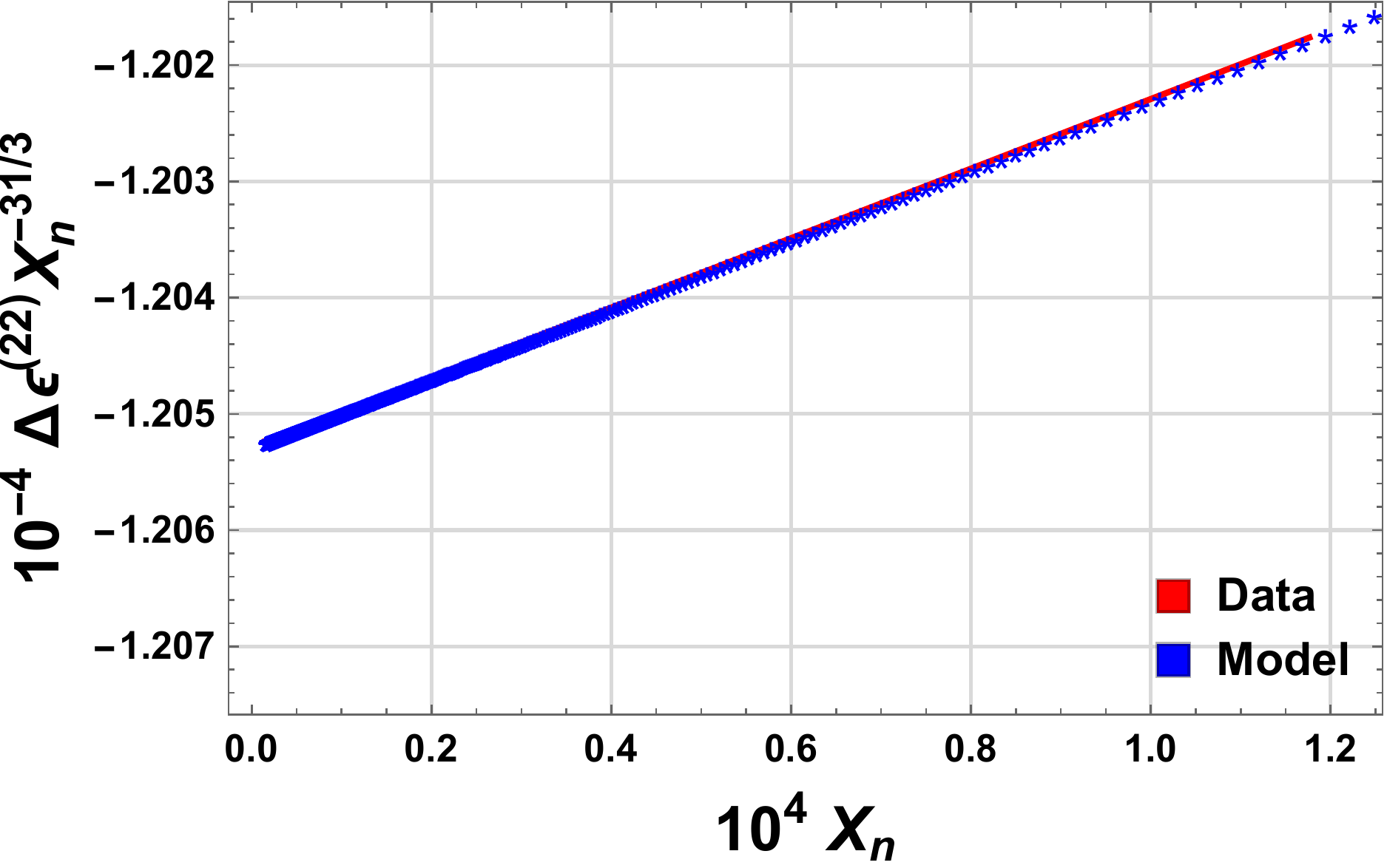}
\caption{The coefficients $B_{22}$ and $B_{24}$ are obtained by a linear fit to the above data.  We plot $B_{22} + B_{24} X_n$ to show how closely this model matches the data.}
\label{fig:NWKB}
\end{figure}

We can use our highly accurate energies to extract higher order coefficients.
We define the deviation from the $2m$-th order WKB approximation as
\begin{equation}
\D \eps^{(2m)}_n = \eps_n - \eps_{WKB,n}^{(2m-2)},
\end{equation}
which, according to Eq. \ref{WKBexpl}, has the asymptotic form
\begin{equation}
\D \eps^{(2m)}_n = B_{2m} X_n^{m - 2/3} + B_{2m + 2} X_n^{m - 2/3 + 1} + ...,
\end{equation}
where
\begin{equation}
X_n = \left(n + \frac{1}{2}\right)^{-2},
\end{equation}
yielding
\begin{equation}
\D \eps^{(2m)}_n X_n^{2/3 - m} = B_{2m} + B_{2m + 2} X_n + B_{2m + 4} X_n^2 + ...
\end{equation}
Thus by calculating accurate energies, multiplying them by
$X_n^{2/3 - m}$, and fitting to a line, we confirm the WKB coefficients up to twentieth order
and find the next two coefficients numerically, as shown in Fig. \ref{fig:NWKB}.  
Our most accurate approximations to $B_{22}$ and $B_{24}$ 
were calculated using [3/3000] to be
$B_{22} = -1.2052792 \times 10^4$ and $B_{24} = 2.98 \times 10^5$, which are accurate to the number of digits shown.  To speed up the calculation we took advantage of parity and calculated the odd and even energies separately using the ParallelTable function in Mathematica \cite{W20}.

\sec{Conclusions}

We have used Blinder's method to extract many quantities from the general quartic
oscillator to many digits \cite{B19}.  We have considered many distinct limits and scenarios
where these benchmark results might be useful.   
We have covered energetics of eigenstates,
the virial theorem, tunneling between wells, the sextic oscillator, analytic forms
in a few basis functions, error dependence on choice of $\omega$, perturbation theory in
the quadratic term, asymmetric wells, and asymptotic analysis of WKB results
for the pure quartic case.
In all cases, we have provided
preliminary analysis and compared with the exact results.  
Some of this work should also prove useful for pedagogy.
This would include both the use of Mathematica to generate accurate
results and the derivations of various results in this context.  Users who wish to replicate our results can start with Ref. \cite{B19} and modify the Hamiltonian with a quartic potential using the matrix elements in Eq. \ref{MatrixElem}. But the two examples of asymptotic techniques are beyond most standard
curricula, and unfamiliar to most theorists.  Refs. \cite{B20b,BO99,BB19,BB20} provide a
pedagogical introduction to such methods.

\sec{Funding Information}
This research was supported by NSF (CHE 1856165).

\pagebreak
\appendix
\sec{Derivation of asymptotic splitting formula}
We now explain how to derive Eq. \ref{AWKB}, the asymptotic approximation to $\D\eps = \eps_1 - \eps_0$ in the limit $\lam \rightarrow \infty$.  

We introduce the shorthand $\eta = (2/\lam)^{3/4}$ so $\lam \rightarrow \infty \implies \eta \rightarrow 0_+$.  In terms of $\eta$ Eq. \ref{CFI} of the main text becomes
\begin{equation}
\label{PhiEta}
\phi(\eta) = \frac{4}{\eta^2} \int_{0}^{\sqrt{1 - \eta}} dx \sqrt{(1 - x^2)^2 - \eta^2}.
\end{equation}
In the limit $\eta \rightarrow 0_+$,
\begin{equation}
\label{ZO}
\phi^{(0)}(\eta) = \frac{4}{\eta^2} \int_{0}^{1} dx (1 - x^2) = \frac{(2 \lam)^{3/2}}{3}.
\end{equation}
We evaluate Eq. \ref{PhiEta}:
\begin{equation}
\label{PhiAnal}
\phi(\eta) = \frac{8}{3 \eta^2} \sqrt{1 + \eta} F(\eta),
\end{equation}
where $F(\eta) = \mathcal{E}(y) - \eta \mathcal{K}(y)$, $y =  (1 - \eta)/(1 + \eta)$ and
\begin{equation}
\mathcal{K}(x) = \int_{0}^{\pi/2} \frac{d\t}{f(x,\theta)}, \hspace{1mm} \mathcal{E}(x) = \int_{0}^{\pi/2} d\t f(x,\theta),
\end{equation}
with $f(x,\theta) = \sqrt{1-x \sin ^2 \t}$ \cite{NIST2}.  The following expansion will prove useful shortly:
\begin{equation}
\label{AsymExpan}
F(\eta) = 1 - \frac{\eta}{2} + \frac{3}{16} \eta^2 (1- 6 \ln 2 + 2 \ln \eta) + \mathcal{O}(\eta^3),
\end{equation}
as $\eta \rightarrow 0_+$ \cite{NIST1}.  Inserting Eq. \ref{AsymExpan} into Eq. \ref{PhiAnal} and expanding around $\eta = 0$, yields
\begin{equation}
\phi^{(2)}(\lam) = \frac{(2 \lam)^{3/2}}{3} - \frac{3}{4} \ln \lam - \frac{1}{4} (2 + 9 \ln2).
\end{equation} 
The above equation combined with Eq. \ref{13} leads to the final result, Eq. \ref{AWKB} of the main text.

\end{document}



\sf 
\coloredtitle{Uncommonly accurate energies for the general quartic oscillator: Supplemental Information}
\author{\color{CITECOL} Pavel Okun, Kieron Burke}
\affiliation{Department of Chemistry,
University of California, Irvine, CA 92697,  USA}
\affiliation{Departments of Physics and Astronomy and of Chemistry}
\date{\today}


\maketitle
\def\floor#1{{\lfloor}#1{\rfloor}}
\def\sm#1{{\langle}#1{\rangle}}
\def\dis{_{disc}}
\newcommand{\Z}{\mathbb{Z}}
\newcommand{\R}{\mathbb{R}}
\def\w{^{(0)}}
\def\w{^{\rm WKB}}
\def\II{^{\rm II}}
\def\hd#1{\noindent{\bf\textcolor{red} {#1:}}}
\def\hb#1{\noindent{\bf\textcolor{blue} {#1:}}}
\def\eps{\epsilon}
\def\ew{\epsilon\w}
\def\ej{\epsilon_j}
\def\upet{^{(\eta)}}
\def\ejeta{\ej\upet}
\def\tjeta{\tj\upet}
\def\bej{{\bar \epsilon}_j}
\def\ewj{\epsilon\w_j}
\def\tj{t_j}
\def\vj{v_j}
\def\F{_{\sss F}}
\def\xt{x_{\sss T}}
\def\sc{^{\rm sc}}
\def\al{\alpha}
\def\ae{\al_e}
\def\bj{\bar j}
\def\bz{\bar\zeta}
\def\eq#1{Eq.\, (\ref{#1})}
\def\cN{{\cal N}}

\graphicspath{{./F/}}
\def\lam{\lambda}
\def\G{\Gamma}
\def\g{\gamma}
\def\eps{\epsilon}
\def\om{\omega}
\def\D{\Delta}
\def\a{\alpha}
\def\t{\theta}

\begin{table*}[!ht]
\renewcommand\thetable{S1} 
\scalebox{0.7}{
\begin{tabular}{|c|r|r|r|}
\hline
n & \multicolumn{1}{c|}{$\lambda = - 1$} & \multicolumn{1}{c|}{$\lambda = 0$} & \multicolumn{1}{c|}{$\lambda  = 1/2$} \\
\hline
0 & 0.6209270298257486608580357329871206982000 & 0.4208049744754477632073387069472240693432 & 0.2969675302972151588212005493833645945082 \\
1 & 2.0259661641666569970850703427960975727209 & 1.5079012411604822141183717271129771258040 & 1.2083664545337553244675128781490877007052 \\
2 & 3.6984503193780828535724670322135994784906 & 2.9587956874793210438093895440143405907473 & 2.5549656461358335669363203892249780748117 \\
3 & 5.5575771385568190043356690869633769327987 & 4.6212203186659158345199462032646287773490 & 4.1185132911873791789827356847060866472967 \\
4 & 7.5684228735599952483040236007700297795874 & 6.4535099323116714688390801378688895232907 & 5.8615290772456843939267326829532388801763 \\
5 & 9.7091478766133491585283420979384357791199 & 8.4284538781251624410812759084851264487491 & 7.7535437904317968231655053084272429114240 \\
6 & 11.9645436206307893693476637072524067567336 & 10.5278307660214238235047658353664143415435 & 9.7748772610230392921405211594194456457452 \\
7 & 14.3232651989715130212955084127970016404927 & 12.7383369432737903427451672925556149750443 & 11.9112456942270964580082777754683535757181 \\
8 & 16.7764527862477299481795270148716610404336 & 15.0497529310262142774202965577869187310261 & 14.1517497832384947562615037963742371519444 \\
9 & 19.3169543039861738015603041224460630143480 & 17.4539341576043532672221679689954879136362 & 16.4877481831099757519730264219436557939311 \\
10 & 21.9388493623736883118412614839274832048677 & 19.9442078219410104858697265744541155274941 & 18.9121913414809377170587410723808518620447 \\
11 & 24.6371393784850827234483399467490277731788 & 22.5149877684426714142542318593734791245252 & 21.4191988384016334155749968573601998934125 \\
12 & 27.4075361599007733167185785640547638285198 & 25.1615158711766884276146901754352125654444 & 24.0037771856458996617329273486803658710224 \\
13 & 30.2463121638354802813346902742580711908208 & 27.8796812841877656582298671760453383565028 & 26.6616235644061192930466358802950389698173 \\
14 & 33.1501911449899797647855437419461171831573 & 30.6658899335127314628439893169538951067637 & 29.3889847264493968907502226461423327368738 \\
15 & 36.1162662359316612997904513472041145489241 & 33.5169677028528987634937196368642936314747 & 32.1825527201796215069940450096463713573417 \\
16 & 39.1419372382007939004180627624271107209340 & 36.4300869553098956349617834567278369397641 & 35.0393860174731882492006918400771100061893 \\
17 & 42.2248617205992722354303466637426801349502 & 39.4027096653309522764323911075252220491304 & 37.9568486543314530223356956562434008727611 \\
18 & 45.3629162655749501221305299308397198651343 & 42.4325426544203398670711200468442512975238 & 40.9325624549159822838446125803113988960005 \\
19 & 48.5541653210865955322219587604320298774269 & 45.5175018286087836091156740676820857337808 & 43.9643689568979728052340425260495584501329 \\
\hline
n & \multicolumn{1}{c|}{$\lambda  = 1$} & \multicolumn{1}{c|}{$\lambda  = \lambda _c$} & \multicolumn{1}{c|}{$\lambda  = 2$} \\
\hline
0 & 0.1472351400900356499691248977564660173258 & -0.0000000000000000000000000000000000000000 & -0.2995213674158870148305588119678013665817\\
1 & 0.8722611979074849376077725041342821484402 & 0.5716588679960300376877118960786601938386 & 0.0463710822278322225551254617719892396217 \\
2 & 2.1279787487595877045538224256388430610159 & 1.7736948620145091772993229646471235886632 & 1.2279728124733779318126911954478247721008 \\
3 & 3.5910891429085531474677426173361466615617 & 3.1528326580224174810921481032323232487809 & 2.4598414289848618278734212356647449784682 \\
4 & 5.2451344076798266430684130776728360862553 & 4.7363694975061665860330732922691834287200 & 3.9382619668923261066022151910142706869605 \\
5 & 7.0543309765744167117016069821477629262677 & 6.4796199486673050896924411810029763381620 & 5.5812919464200356086062138581050168522645 \\
6 & 8.9977152866347239207805152219715373495070 & 8.3609827938705988454021151460423261599059 & 7.3688888891530412336227399929097153100735 \\
7 & 11.0600197362083473113526351645076563394713 & 10.3643546811859909563918554922259918339922 & 9.2832226299648908948309227497618828073426 \\
8 & 13.2296716965972822290960283911125387262757 & 12.4776265015870503319630402009506998689629 & 11.3113496812053433661018404740221218764899 \\
9 & 15.4975367732762450862155724272143845256796 & 14.6912708225153021462064097897940203073039 & 13.4431253746146181822961187485295612840726 \\
10 & 17.8561912029006493999576787885895573682737 & 16.9975664549772044605549383045427198392948 & 15.6703699687296553967537802445771571591652 \\
11 & 20.2994620601279773169062848753786930728930 & 19.3901082427550754621639190325828280149029 & 17.9863231152929266903954684521570229693708 \\
12 & 22.8221217502876403219503896506332221428776 & 21.8634828235108640404426731524651415625608 & 20.3852887904825910828332734073819599201359 \\
13 & 25.4196764022940486530210893302948537481785 & 24.4130447540232097527558401520261731285525 & 22.8623916753093559460983299855869625178436 \\
14 & 28.0882142843477951830862545628663297491065 & 27.0347565735505259647623769404174440869892 & 25.4134040381637348805306139183152990003897 \\
15 & 30.8242941384532552138898486648626838905608 & 29.7250712630490587442470044304579531460627 & 28.0346190393585560070298786629180414903382 \\
16 & 33.6248609575586778848715479117742888261600 & 32.4808437755606569236339263329500479754022 & 30.7227557029688189378854855857924222071042 \\
17 & 36.4871811657174959586755707609740381069210 & 35.2992630749155671032364812414301202000714 & 33.4748861345856088549750625423586830442841 \\
18 & 39.4087918519281065606950050769764515801679 & 38.1777989988783344057067262022906053148024 & 36.2883787630669702226446175430707502923971 \\
19 & 42.3874603986599763607484603391513404125215 & 41.1141600671899623565300558968349506123263 & 39.1608533771374500623766820576210674574060 \\
\hline
n & \multicolumn{1}{c|}{$\lambda  = 4$} & \multicolumn{1}{c|}{$\lambda  = 8$} & \multicolumn{1}{c|}{$\lambda  = 16$} \\
\hline
0 & -2.6614480668299258763230139306673458375429 & -14.0324444018052994432537497688658953282978 & -61.1873976097239347040519514878376408475110 \\
1 & -2.6517317169216879626798264247619112670590 & -14.0324443753846106821056639056976599894887 & -61.1873976097239347040519507432046834481584 \\
2 & -0.5102930381327473696110504044071024379800 & -10.2374181022536577826397549201626978893469 & -55.6270328066566950481384614788601681896687 \\
3 & -0.1807894330622582818165542233132232566996 & -10.2374126036757771174222559086720970290432 & -55.6270328066566950481379569833156389101676 \\
4 & 1.1695143365947340664211154850717789208582 & -6.6843927138660009112120807253579250040901 & -50.1677907702224152602395145875345725833160 \\
5 & 2.3643918923460492284769310044183117645863 & -6.6839252868713098232048203999239475485623 & -50.1677907702224152600774529069586335173511 \\
6 & 3.8357948326004093266558747356504382940929 & -3.4521491483210872751782055732716385290964 & -44.8154388106509680002392073351225222773039 \\
7 & 5.4430045231550815592562510060115461795846 & -3.4323617805610231231123996788750204987627 & -44.8154388106509679674606247338249029670626 \\
8 & 7.1832349741044744228694202343455748298609 & -0.8436487231434934433791826110408175496539 & -39.5766864040676713567900013956791558340159 \\
9 & 9.0398481121463461930986519520826284005100 & -0.5042404268290690707095844038061155129498 & -39.5766864040676666831144013830302553496714 \\
10 & 11.0025188337220857851315919678616051381955 & 1.1615932648322248170487151971678725823886 & -34.4594623366015919928440409675286441966089 \\
11 & 13.0627969100066585323849398998537049950316 & 2.3918554218704864477402234706269532999536 & -34.4594623366010937848302698406235040845512 \\
12 & 15.2137826575078115399041761005777164302396 & 3.9821840692096918952249865418185404041377 & -29.4733189381534883371911762187717947850975 \\
13 & 17.4496779348669813230194759912660986824602 & 5.6581105244767300215025449175194901075413 & -29.4733189381124209160128178684499808471623 \\
14 & 19.7655245277330684324543398617660707059238 & 7.4555769375507348817071475302379978220562 & -24.6300488773159008572118670474641144980755 \\
15 & 22.1570215163914948104597068950418213904794 & 9.3507619349088529032629887250749107416157 & -24.6300488746474233947975266198680141507504 \\
16 & 24.6203946769103286322488223100432640789732 & 11.3367520808888588886234641628154072096683 & -19.9446824518702288293797651016978129099597 \\
17 & 27.1522997333340460883361104535392557117227 & 13.4065360556998380563145435100523133286453 & -19.9446823139694465593687674516291436652453 \\
18 & 29.7497488149246926808474288802922162419641 & 15.5547993768313120592069318313123633281790 & -15.4372324690911890964651244057502958813076 \\
19 & 32.4100533460587587533833010330720289301049 & 17.7770953113984742550117244948104017128921 & -15.4372268038006350393996616788105217144393 \\
\hline
\end{tabular}}
\caption{Exact energies given to 40 digits.}
\label{tab:S1}
\end{table*}
\begin{table*}[!ht]
\renewcommand\thetable{S2} 
\scalebox{0.5}{
$\begin{array}{|c|r|r|r|r|}
\hline
\text{n} & \multicolumn{1}{c|}{\lambda = 0} & \multicolumn{1}{c|}{\lambda  = \lambda _c} & \multicolumn{1}{c|}{\lambda = 4} & \multicolumn{1}{c|}{\lambda  = 8} \\
\hline
0 & -9.7588252368918194861531575958132027216984\times 10^{-01} & -8.9891335378216959429748135104379263992213\times 10^{-01} & 2.6366736960904910080338141237575471683707\times
10^{-01} & -1.2284345041748225948485885350411922442056\times 10^{-02} \\
2 & -2.1710763920798390016468830973743157043024\times 10^{-01} & -4.1893751204555503433953849279983557892984\times 10^{-01} & 5.9525613945048726688710745173110956505724\times
10^{-01} & -9.5000561727618961112074124321497315955482\times 10^{-02} \\
4 & -2.2261890633146942297495023994167441677613\times 10^{-02} & -1.2754928931029004588560371047300404192716\times 10^{-01} & 6.3755092396869292512676919553577799959564\times
10^{-01} & -2.8895050102284264633264692648801269419409\times 10^{-01} \\
6 & 4.5591557787546429705804003601242237486974\times 10^{-03} & -1.2197224794254465950455087492124630583470\times 10^{-02} & 3.9252576753721657029342146416395277347336\times
10^{-01} & -5.1751005800289649992548077856145587327207\times 10^{-01} \\
8 & 1.0525477149971613425352573771627949721315\times 10^{-03} & 5.2238031507692678188583838891905615620312\times 10^{-03} & 1.2426035934929888629492095903427597771632\times
10^{-01} & -6.0644761277763321830852528870076142585510\times 10^{-01} \\
10 & -2.9103720020779682673437831128086393462750\times 10^{-04} & 1.0415337942402997588758229945604233264700\times 10^{-03} & 4.5218782880604521880565599178279919182618\times
10^{-4} & -4.7036546674350963606011473974687141511015\times 10^{-01} \\
12 & -4.4315556015676489960413529539763509157754\times 10^{-05} & -4.1366570716572072358700463224198391959064\times 10^{-04} & -1.2548603768400798394430043971321101311319\times
10^{-02} & -2.2049343759930470715570650167306954763487\times 10^{-01} \\
14 & 2.7385266752839770313829213703535806999331\times 10^{-05} & -6.1195308111003926284232317755775636245208\times 10^{-05} & -1.2175962201115102957708710468681604815135\times
10^{-03} & -3.4987244075933380837161542327009269932314\times 10^{-02} \\
16 & -7.9795051736373767819752776251262980100650\times 10^{-07} & 4.5775803526838129469539954283393712137855\times 10^{-05} & 1.4642210813276674071071389312089900357357\times
10^{-3} & 2.2679742539034376005681249911459389846413\times 10^{-02} \\
18 & -2.3255573994102993778937068556143188251051\times 10^{-06} & -7.3856649473612454176961564069524459383129\times 10^{-07} & 1.0184874757133244100699395435973076163382\times
10^{-04} & 1.1288188541125057476075179217375706247192\times 10^{-02} \\
20 & 6.0405679973221260534531213171483279655685\times 10^{-07} & -4.8304782500867527035471627264981826180864\times 10^{-06} & -2.0291462118463796953113826820587548801465\times
10^{-04} & -2.0872973146669747780267890506998273648440\times 10^{-03} \\
22 & 8.8908424687998948857843126583955387567507\times 10^{-08} & 1.1391556241250729079831506137187730517943\times 10^{-06} & 1.1971425631666171747637084065481945348537\times
10^{-05} & -2.1611523559764022566630973228900028755257\times 10^{-03} \\
24 & -9.1179059386670698706017409888235839985727\times 10^{-08} & 3.0282068165362269340638867139270623603811\times 10^{-07} & 2.7253159877626026184652046424695602473164\times
10^{-05} & 3.2641654407826534784672309359832959707374\times 10^{-04} \\
26 & 1.7156823987605295867673461189168296473864\times 10^{-08} & -2.2650031555948765329240285593413319168973\times 10^{-07} & -7.2642193953197722988401484963607687076995\times
10^{-06} & 3.8984304686928272732587820030098062386320\times 10^{-04} \\
28 & 5.4977888943382273070125919731874506339984\times 10^{-09} & 2.7279610628244718735068998902616798396469\times 10^{-08} & -2.5161581959913066855075598093168741264401\times
10^{-06} & -9.2021897613604714608495654249965283354249\times 10^{-05} \\
30 & -4.0388146366596254001705595403162371603752\times 10^{-09} & 2.2222391095621187887681361282250184444194\times 10^{-08} & 1.8055794560507756604150809656607793751033\times
10^{-06} & -6.4247385262771323412930234669178431327181\times 10^{-05} \\
32 & 7.3684474749356136042733998470002895607617\times 10^{-10} & -1.1376224207158825654152478462343844353211\times 10^{-08} & -1.0521733326394967485197840802108532222554\times
10^{-07} & 2.8212657952347910390642783268812842942531\times 10^{-05} \\
34 & 2.6740668465809183846623078834803341489088\times 10^{-10} & 8.3821985504034972912328250210051901968154\times 10^{-10} & -2.6890442021974607708050558892611841751208\times
10^{-07} & 7.4643427755618958336463449927503457702870\times 10^{-06} \\
36 & -2.0798818500045227492292783430143920452527\times 10^{-10} & 1.3872207993743820800331787791897246292561\times 10^{-09} & 1.0686500064662085219943334421592023503579\times
10^{-07} & -7.4024081291889180641596440716166115949559\times 10^{-06} \\
38 & 4.6750232067405078860334127973060710523926\times 10^{-11} & -6.7312113379401529404366647382616103264719\times 10^{-10} & 9.4134816393796881142710890769561810455547\times
10^{-09} & 3.1337303496364493150989674436562304320167\times 10^{-07} \\
40 & 1.0834231779096182435629173430718530734598\times 10^{-11} & 5.6445347676440645232387553843689740190894\times 10^{-11} & -2.2497111453004683246952298066210992613525\times
10^{-08} & 1.4657200161976769829131116181528658839241\times 10^{-06} \\
42 & -1.1804012724863719142258770641850184199093\times 10^{-11} & 8.4513339069423978325585966951633469717546\times 10^{-11} & 7.0112465042242994476359889505080886579693\times
10^{-09} & -5.0792512169456612192938704249127082757261\times 10^{-07} \\
44 & 3.6446703035983857136480428392440509048217\times 10^{-12} & -4.6250120316613473914740404393564172755806\times 10^{-11} & 1.3175193361307386595255334954723142852777\times
10^{-09} & -1.5139847616210827108984068392399353334543\times 10^{-07} \\
46 & 1.7820180084804597938509676321954224932726\times 10^{-13} & 6.6607539904795598464437339140823074175625\times 10^{-12} & -1.8688186122189692243402082161711037482029\times
10^{-09} & 1.6292818459599496379728276970632111556577\times 10^{-07} \\
48 & -6.7323056188590048204487590458721334061366\times 10^{-13} & 4.8271453135694179799755446230637597320208\times 10^{-12} & 5.6143648258895084734587229767659874172269\times
10^{-10} & -2.5911612501156841882043765253158432835112\times 10^{-08} \\
50 & 2.9833397203797698473932496978475949436735\times 10^{-13} & -3.4579657783968680919121203317145856198250\times 10^{-12} & 1.0958824100429981692906327559738487299334\times
10^{-10} & -2.6449082975155940565903578730768763489912\times 10^{-08} \\
52 & -3.5512503186487994528766855486751303002979\times 10^{-14} & 8.3026331317981105849033659810559746759409\times 10^{-13} & -1.6087404675025286880227701545294437871295\times
10^{-10} & 1.6420563550529854102467275273828539016937\times 10^{-08} \\
54 & -3.2911846474186238511973288988143153936309\times 10^{-14} & 2.0903546439867351150332746937816349239262\times 10^{-13} & 5.4626174912812805360297127362063299676167\times
10^{-11} & -8.0617048897031989578112472742015275930887\times 10^{-10} \\
56 & 2.3231861879176994112022830553338399266214\times 10^{-14} & -2.5848182361169422886679199715570037936389\times 10^{-13} & 5.8132501440743133664051797346823869967792\times
10^{-12} & -3.4316970166708484129842490665767423867353\times 10^{-09} \\
58 & -6.4593999639108647230639217868283910755797\times 10^{-15} & 9.5537482294158070174328908989056422660877\times 10^{-14} & -1.4051708950879115906795572038361989532930\times
10^{-11} & 1.7192318007587832279717516991927567901671\times 10^{-09} \\
60 & -6.9325243632049375843603880680769856547413\times 10^{-16} & -1.9045609695393518011981035090993673896733\times 10^{-15} & 5.9669769807247903069744423578090572314474\times
10^{-12} & 1.3316873580855187750074738375862056961619\times 10^{-11} \\
\hline
\text{Norm. Error} & 2.981756840031728465318268238576759389575\times 10^{-30} & 3.9882215351120094379193997210890776195050\times 10^{-28} & 1.8720770271652866139815322043394623283636\times
10^{-24} & 2.1272947859720467040290652122039873518905\times 10^{-19} \\
\hline
\end{array}$}
\caption{The expansion coefficients of the ground state for several values of $\lam$ calculated with [2/200].  The normalization error is defined as $\text{Norm. Error} = 1 - \sum_{n=0}^{30}c_{2n}^2$.  All entries are accurate for all digits shown.}
\label{tab:S2}
\end{table*}
\begin{table*}[!ht]
\renewcommand\thetable{S3} 
\scalebox{0.7}{
$\begin{array}{|c|r|r|r|r|}
\hline
\lam  & \multicolumn{1}{c|}{\expval{p^2}} & \multicolumn{1}{c|}{\expval{x^2}} & \multicolumn{1}{c|}{\expval{x^4}}& \multicolumn{1}{c|}{\expval{p^2} + \lam \expval{x^2} - \expval{x^4}} \\
\hline
 -1 & 0.70962262271004694816315661942639984135924 & 0.35484025117285379894267307366928326872234 & 0.35478237153719314922048354575711657263690 & -1.0\times
10^{-69} \\
0 & 0.56107329930059701760978494259629875912425 & 0.45611995574755266412568999336488624453373 & 0.56107329930059701760978494259629875912425 & -3.8\times
10^{-68} \\
\frac{1}{2} & 0.48595283076997202057003770348474894873410 & 0.53997674224211085285062182584157693633932 & 0.75594120189102744699534861640553741690376
& -3.2\times 10^{-66} \\
1 & 0.41875308378513724691880827621103099209885 & 0.66731869099526914087992523760722890699352 & 1.08607177478040638779873351381825989909236 & 9.6\times
10^{-67} \\
\lambda _c & 0.38288731030800749229582749908269219838526 & 0.82149466176780355559023407522345874736335 & 1.53154924123202996918330999633076879354117
& 0.0\times 10^{-40} \\
2 & 0.40538382521367762484168184235523299173570 & 1.20711847265229046692364038746845222076689 & 2.81962077051825855868896261729213743326949 & 1.2\times
10^{-63} \\
4 & 1.22302810893072218747770900747062384312444 & 3.57871914852796751693129568627031371988625 & 15.53790470304259225520289175255187872266944 & 9.7\times
10^{-60} \\
8 & 1.93380805077592808747294114028235780328191 & 7.74140021994362275442922781203883184037962 & 63.86500981032491012290676363659301252631885 & -1.6\times
10^{-51} \\
\hline
\end{array}$}
\caption{Testing the virial theorem on the ground state in the basis [2/200] for various $\lam$.}
\label{tab:S3}
\end{table*}
\begin{table*}[!ht]
\renewcommand\thetable{S4} 
\scalebox{0.6}{
$\begin{array}{|r|r|r|}
\hline
\multicolumn{1}{|c|}{\lam}  & \multicolumn{1}{c|}{\epsilon _{\text{ZP}}} & \multicolumn{1}{c|}{\Delta \epsilon} \\
\hline
- 1.0 & 0.62092702982574866085803573298712069820002 & 1.4050391343409083362270346098089768745208\times 10^{-0} \\
- 0.9 & 0.60299632226964628844798231393579064116412 & 1.3751166990600236139114185321635452644040\times 10^{-0} \\
- 0.8 & 0.58467407401334478858446698193363989988428 & 1.3448018746401551940280748998539462115265\times 10^{-0}  \\
- 0.7 & 0.56593768667208973311389818692106524345099 & 1.3140853140631193365254355121211365308478\times 10^{-0}  \\
- 0.6 & 0.54676256728408263864067410130046858482152 & 1.2829578098203442558192144577403628641190\times 10^{-0} \\
- 0.5 & 0.52712189910631718948160847475225511600155 & 1.2514103967290150110732296148266232837839\times 10^{-0} \\
- 0.4 & 0.50698638094818818887295400349059963960776 & 1.2194344778284042769140563256722639977853\times 10^{-0}  \\
- 0.3 & 0.48632393021779125714651265825852295399743 & 1.1870219779052869196890031652748767144085\times 10^{-0} \\
- 0.2 & 0.46509934409245165218048652179741992023225 & 1.1541655300384905427168316854046433487992\times 10^{-0} \\
- 0.1 & 0.44327391235674714731202844741697456611339 & 1.1208587015272327908233317548892182205668\times 10^{-0} \\
 0.0 & 0.42080497447544776320733870694722406934319 & 1.0870962666850344509110330201657530564608\times 10^{-0} \\
0.1 & 0.40014541239049915015655202895057004496732 & 1.0528745352429691823424756146892005253677\times 10^{-0} \\
0.2 & 0.38374306936871792774843000399336180905692 & 1.0181917465029299214242205320914791197489\times 10^{-0} \\
0.3 & 0.37154008401984630588960982686231893532549 & 9.8304854088319486418257971399998175236849\times 10^{-1} \\
0.4 & 0.36347212742599817298561789952511970382945 & 9.4744852204198535827549906077458178738227\times 10^{-1} \\
0.5 & 0.35946753029721515882120054938336459450816 & 9.1139892423654016564631232876572310619703\times 10^{-1} \\
0.6 & 0.35944628639370443462820796244243875991543 & 8.7491140078599896968941055709766129850312\times 10^{-1} \\
0.7 & 0.36331891844401198220440581192726810360345 & 8.3800295015483345935974872247581416559151\times 10^{-1} \\
0.8 & 0.37098519390821073853744677368109326090619 & 8.0069699579858230964730291789504576910103\times 10^{-1} \\
0.9 & 0.38233268087228305029630450855134894431594 & 7.6302463383245331359843389987475231600721\times 10^{-1} \\
1.0 & 0.39723514009003564996912489775646601732576 & 7.2502605781744928763864760637781613111446\times 10^{-1} \\
1.1 & 0.41555075905284850269679293862634027477515 & 6.8675216115574772785403996013260132180676\times 10^{-1} \\
1.2 & 0.43712024972864869131921263087196878733833 & 6.4826630296005290613509983024275992615350\times 10^{-1} \\
1.3 & 0.46176485545137583832221671072501128088771 & 6.0964620058917679728128761061240051872668\times 10^{-1} \\
1.4 & 0.48928434681380932526184346589578065540815 & 5.7098587881161888311010534076421846183981\times 10^{-1} \\
1.5 & 0.51945513362201026928354780954310547217758 & 5.3239755938515668689983574067719278771953\times 10^{-1} \\
1.6 & 0.55202868127393456146258716772935164712692 & 4.9401331434894835185037107761821939227067\times 10^{-1} \\
1.7 & 0.58673049398989763593297474697028223045688 & 4.5598623271756178977257728951253039247290\times 10^{-1} \\
1.8 & 0.62326000782266187450311983649619680518098 & 4.1849076971005900299967590231490230188190\times 10^{-1} \\
1.9 & 0.66129180897661221034711073671549879025925 & 3.8172187430012034691909211233590839964834\times 10^{-1} \\
2.0 & 0.70047863258411298516944118803219863341833 & 3.4589244964371923738568427373979060620339\times 10^{-1} \\
2.1 & 0.74045656751793543556297749966876954520104 & 3.1122872787829673232852391573241979544497\times 10^{-1} \\
2.2 & 0.78085275210538198578548373047451121421126 & 2.7796327774199604817854862992449978000166\times 10^{-1} \\
2.3 & 0.82129556061745322182658509210821302751409 & 2.4632564354606722765049741839955396695579\times 10^{-1} \\
2.4 & 0.86142685094135521447815747208424095729840 & 2.1653103753875066326339072313185707851793\times 10^{-1} \\
2.5 & 0.90091532884444051424031271067679910450119 & 1.8876801593314337098477735325258520829547\times 10^{-1} \\
2.6 & 0.93946961245818474504934597615144484627339 & 1.6318653348952627884979765105200671202821\times 10^{-1} \\
2.7 & 0.97684932483348441178014034279342247846547 & 1.3988802033057390329223388125239158098869\times 10^{-1} \\
2.8 & 1.01287265091553830734575456520266331995205 & 1.1891901003987891328858434406118046691496\times 10^{-1} \\
2.9 & 1.04741930707915501699739968720119489835874 & 1.0026933176210712058637366222793647465800\times 10^{-1} \\
3.0 & 1.08042866946106801495684311318592662778675 & 8.3875078659544046104422960854141328549576\times 10^{-2} \\
3.1 & 1.11189365349778014894835720613139852497348 & 6.9625719425334859727642078483686106686743\times 10^{-2} \\
3.2 & 1.14185157804341337706730782782970238305219 & 5.7374082131274903868486094877790640131769\times 10^{-2} \\
3.3 & 1.17037352780656492384274271631107190848152 & 4.6947665787295806496556615913278874617551\times 10^{-2} \\
3.4 & 1.19755364117134058441182535135168528617026 & 3.8159830575992726850779420745286423643317\times 10^{-2} \\
3.5 & 1.22349940926910679491009773791065538911109 & 3.0819769492830925530425153918943404031434\times 10^{-2} \\
3.6 & 1.24832363249037097977123058441409539280664 & 2.4740615018333114186125336647150711310158\times 10^{-2} \\
3.7 & 1.27213827316211688448713837379474356017189 & 1.9745452707883447866858594904011205567527\times 10^{-2} \\
3.8 & 1.29505014056664534673242524491780119766838 & 1.5671325524158315378419332481735730221177\times 10^{-2} \\
3.9 & 1.31715816401351639865255691713621587598623 & 1.2371500431668495730106131763125365198563\times 10^{-2} \\
4.0 & 1.33855193317007412367698606933265416245708 & 9.7163499082379136431875059054345704839343\times 10^{-3} \\
4.1 & 1.35931118040280730318426137227983642479201 & 7.5932074094905055157187782958462932145716\times 10^{-3} \\
4.2 & 1.37950591636752981130618879703964781048167 & 5.9055187708652852174274643546950108377511\times 10^{-3} \\
4.3 & 1.39919698383856090730912395360910305136533 & 4.5715551589515568970514585700712973116190\times 10^{-3} \\
4.4 & 1.41843685070354771824883434650023917688893 & 3.5228931453146243630821718978213512658769\times 10^{-3} \\
4.5 & 1.43727051315036263002223155068628293974643 & 2.7028124016803728740771084600651645279280\times 10^{-3} \\
4.6 & 1.45573642112777618127626866164577488169062 & 2.0647151777248071188829247572436858948968\times 10^{-3} \\
4.7 & 1.47386736978693179203952476202087894636313 & 1.5706349720270040415072252652820979744935\times 10^{-3} \\
4.8 & 1.49169132381879049275175607071709374289339 & 1.1898739683345029873121740413665681732361\times 10^{-3} \\
4.9 & 1.50923215792196271279341308066738162102173 & 8.9778856323548681311254855617471183212927\times 10^{-4} \\
5.0 & 1.52651030762976358522283190644720065113446 & 6.7472818643965346851187809866564543341188\times 10^{-4} \\
5.1 & 1.54354333177388361448261448937392388361459 & 5.0512323352521319469615727925036261857589\times 10^{-4} \\
5.2 & 1.56034639208590120184912433046669370208156 & 3.7671210909198439225958135329624669159761\times 10^{-4} \\
5.3 & 1.57693265770268067535117460755759835559144 & 2.7989414843849389130603193469809827015170\times 10^{-4} \\
5.4 & 1.59331364329207792412816979205470088012467 & 2.0719378468138621419557320861418463764247\times 10^{-4} \\
5.5 & 1.60949948961794038007158775988250964986722 & 1.5282116884112167535087805873088229492202\times 10^{-4} \\
5.6 & 1.62549919494790249397693001850340691739592 & 1.1231509218305786768051396881601758100642\times 10^{-4} \\
5.7 & 1.64132080500163692479578076775771321136715 & 8.2255179762697894570370375023656230888559\times 10^{-5} \\
5.8 & 1.65697156829429409728197859354937942539907 & 6.0031690599786056690293062049561302619933\times 10^{-5} \\
5.9 & 1.67245806284981737365245447915582391395264 & 4.3662713149159382641171716887986553895338\times 10^{-5} \\
6.0 & 1.68778629940407124803045902865463863637959 & 3.1649978415005754030679041765499957469812\times 10^{-5} \\
6.1 & 1.70296180542465691824625723987731026311021 & 2.2865860656074198828686947575121671356759\times 10^{-5} \\
6.2 & 1.71798969356162388143694046242318430452398 & 1.6465359482304028840146632006319000101076\times 10^{-5} \\
6.3 & 1.73287471751801736768578542563039009667578 & 1.1817939893854221014674611055385316725886\times 10^{-5} \\
6.4 & 1.74762131779084577252135709539203256612564 & 8.4550449100027037869708136352867731984463\times 10^{-6} \\
6.5 & 1.76223365927660122682765687239875757169333 & 6.0298942490602010084911548282466379150117\times 10^{-6} \\
6.6 & 1.77671566235342490098061178484950242603425 & 4.2868527685477105919919000260260566479292\times 10^{-6} \\
6.7 & 1.79107102873576958186798938254052947995874 & 3.0382075670596470239716086191584004070007\times 10^{-6} \\
6.8 & 1.80530326313815605719761957100910285216709 & 2.1466473231063383744483106596348708677531\times 10^{-6} \\
6.9 & 1.81941569157387699712524561811115286359124 & 1.5121060000153314738892984391111778847897\times 10^{-6} \\
7.0 & 1.83341147694447754194075549762250906434191 & 1.0619290128318458839696765073793069464257\times 10^{-6} \\
7.1 & 1.84729363243957903660764841461593128264079 & 7.4355559422994055230904848405574087749348\times 10^{-7} \\
7.2 & 1.86106503315804713076418828797034369971556 & 5.1909721677295689603051136596798999038775\times 10^{-7} \\
7.3 & 1.87472842627545056546554138986800765404180 & 3.6133784326113664349515192977479612893014\times 10^{-7} \\
7.4 & 1.88828644001484440732937598243310360737341 & 2.5079537135762514013924411972571341080399\times 10^{-7} \\
7.5 & 1.90174159162451200113577827122101273723257 & 1.7357148773654877924571962961217335992384\times 10^{-7} \\
7.6 & 1.91509629452443286375626516550226550197378 & 1.1978465926873801339823410252260274864434\times 10^{-7} \\
7.7 & 1.92835286475048617341755148853391931921428 & 8.2432561597168695731282577658808280215263\times 10^{-8} \\
7.8 & 1.94151352679979968669036785239354554570038 & 5.6569416417113789099506419453458916137301\times 10^{-8} \\
7.9 & 1.95458041896065251558815819669241541295197 & 3.8713296557554231694188905036612432355486\times 10^{-8} \\
8.0 & 1.96755559819470055674625023113410467170220 & 2.6420688761148085863168235338809066342446\times 10^{-8} \\
\hline
\end{array}$}
\caption{Zero point energy ($\eps_{zp}$) and energy difference between the first excited and ground states ($\D \eps = \eps_1 - \eps_0$).  All calculations made with [2/200].}
\label{tab:S4}
\end{table*}
\begin{table*}[!ht]
\renewcommand\thetable{S5}
\scalebox{0.9}{
$\begin{array}{|c|r|r|r|r|r|r|}
\hline
 & \multicolumn{3}{c|}{\D\text{E}_0 \times 10^{17}} & \multicolumn{3}{c|}{\D\text{E}_{19} \times 10^{2}}\\
\hline
\omega  & \multicolumn{1}{c|}{N_B = 40} & \multicolumn{1}{c|}{N_B = 50} & \multicolumn{1}{c|}{N_B = 60} & \multicolumn{1}{c|}{N_B = 40} & \multicolumn{1}{c|}{N_B = 50} & \multicolumn{1}{c|}{N_B = 60} \\
\hline
1.0 & 2.8316898104\times 10^{+04} & 7.6749053454\times 10^{+01} & 1.1671394475\times 10^{-00} & 9.7444572767\times 10^{+02} & 8.3352481902\times 10^{+01} & 1.2928179048 \times10^{-00} \\
1.1 & 1.4303248620\times 10^{+04} & 4.6384048355\times 10^{+01} & 1.6780669551\times 10^{-01} & 4.5861040726\times 10^{+02} & 5.9258296585\times 10^{-00} & 4.6630523664\times10^{-01} \\
1.2 & 4.1469925551\times 10^{+02} & 9.1180293837\times 10^{-01} & 3.6207836723\times 10^{-03} & 1.5887415607\times 10^{+02} & 1.9298196092\times 10^{-00} & 1.1600275198\times10^{-02} \\
1.3 & 2.8742793086\times 10^{+02} & 5.2200569899\times 10^{-01} & 1.1126424115\times 10^{-03} & 2.9392576233\times 10^{+01} & 7.6207973910\times 10^{-01} &7.3423754374\times 10^{-03} \\
1.4 & 4.2356801829\times 10^{+01} & 2.1693812097\times 10^{-02} & 1.9273478687\times 10^{-05} & 3.3821922647\times 10^{-00} & 3.7361785356\times 10^{-02} & 4.2502441003\times10^{-04} \\
1.5 & 3.4693712760\times 10^{-00} & 7.5543662000\times 10^{-03} & 1.0429735781\times 10^{-05} & 3.6273522076\times 10^{-00} & 2.1806061116\times 10^{-02} & 9.1774434632\times 10^{-05}\\
1.6 & 2.8212055885\times 10^{-00} & 7.6197300492\times 10^{-04} & 1.7298724883\times 10^{-07} & 1.4578570253\times 10^{-00} & 5.4608563291\times 10^{-03} & 1.5285380353\times 10^{-05}\\
1.7 & 1.2974200122\times 10^{-01} & 1.3163775851\times 10^{-04} & 1.3929202224\times 10^{-07} & 1.3615884048\times 10^{-01} & 2.6800092716\times 10^{-04}& 1.0983288804\times 10^{-06} \\
1.8 & 1.1692066597\times 10^{-01} & 3.3406363815\times 10^{-05} & 2.5156704460\times 10^{-09} & 8.3377272370\times 10^{-02} & 3.1402017668\times 10^{-04}& 5.0423268708\times 10^{-07} \\
1.9 & 3.1199725859\times 10^{-02} & 2.6229394016\times 10^{-06} & 2.6472252394\times 10^{-09} & 5.7265477288\times 10^{-02} & 3.5933985680\times 10^{-05}& 1.6545958668\times 10^{-08} \\
2.0 & 2.5175389623\times 10^{-03} & 1.7633090014\times 10^{-06} & 5.7037916005\times 10^{-11} & 7.4920148760\times 10^{-03} & 8.6795822484\times 10^{-06}& 1.7462304043\times 10^{-08} \\
2.1 & 4.1048712013\times 10^{-03} & 6.5050994005\times 10^{-08} & 7.1374851548\times 10^{-11} & 2.8547336030\times 10^{-03} & 5.1625999631\times 10^{-06}& 6.0346848022\times 10^{-10} \\
2.2 & 3.4060054560\times 10^{-04} & 1.1202439689\times 10^{-07} & 1.9648391489\times 10^{-12} & 3.0107456778\times 10^{-03} & 2.5349445848\times 10^{-07}& 6.8792884453\times 10^{-10} \\
2.3 & 3.1908063319\times 10^{-04} & 4.1178041626\times 10^{-09} & 2.7238544179\times 10^{-12} & 5.2796166518\times 10^{-04} & 4.1665006440\times 10^{-07}& 4.8835894139\times 10^{-11} \\
2.4 & 1.6416121890\times 10^{-04} & 8.5785121668\times 10^{-09} & 1.0118763143\times 10^{-13} & 1.7617643954\times 10^{-04} & 8.2250771986\times 10^{-08}& 3.2278123120\times 10^{-11} \\
2.5 & 1.2090347328\times 10^{-05} & 7.7903013537\times 10^{-10} & 1.4704204879\times 10^{-13} & 2.5845945333\times 10^{-04} & 2.0151104609\times 10^{-08}& 5.1712524608\times 10^{-12} \\
\hline
\end{array}$}
\caption{The error of the ground and 19th excited states calculated in bases composed of various values of $\om$ and 40, 50, and 60 harmonic oscillator states.}
\label{tab:S5}
\end{table*}
\begin{table*}[!ht]
\renewcommand\thetable{S6} 
$\begin{array}{|c|r|r|r|r|r|}
\hline
\lambda  & \multicolumn{1}{c|}{\eps_0} & \multicolumn{1}{c|}{\eps_0^{(0)}} & \multicolumn{1}{c|}{\eps_0-\eps_0^{(0)}} & \multicolumn{1}{c|}{\eps_0^{(1)}} & \multicolumn{1}{c|}{\eps_0-\eps_0^{(0)}-\eps_0^{(1)}} \\
 \hline
-1  & 0.6209270298 & 0.5000000000 & 0.1209270298 & 0.1875000000   & -0.0665729702 \\
-2  & 0.7825872853 & 0.7071067812 & 0.0754805041 & 0.0937500000   & -0.0182694959 \\
-3  & 0.9206648830 & 0.8660254038 & 0.0546394792 & 0.0625000000   & -0.0078605208 \\
-4  & 1.0426978264 & 1.0000000000 & 0.0426978264 & 0.0468750000   & -0.0041771736 \\
-5  & 1.1530169674 & 1.1180339887 & 0.0349829786 & 0.0375000000   & -0.0025170214 \\
-6  & 1.2543454649 & 1.2247448714 & 0.0296005935 & 0.0312500000   & -0.0016494065 \\
-7  & 1.3485135506 & 1.3228756555 & 0.0256378951 & 0.0267857143   & -0.0011478192 \\
-8  & 1.4368153589 & 1.4142135624 & 0.0226017966 & 0.0234375000   & -0.0008357034 \\
-9  & 1.5202030526 & 1.5000000000 & 0.0202030526 & 0.0208333333   & -0.0006302807 \\
-10 & 1.5993998650 & 1.5811388301 & 0.0182610350 & 0.0187500000   & -0.0004889650 \\
 \hline
\end{array}$
\caption{Comparing perturbation theory with exact results.  All values are accurate to the number of digits shown.}
\label{tab:S6}
\end{table*}